\documentclass[aps,twocolumn,nofootinbib,showkeys,noeprint,notitlepage,superscriptaddress,floatfix]{revtex4-2}
\usepackage{natbib}
\usepackage{graphicx}
\usepackage{dcolumn}
\usepackage{bm}
\usepackage{amsmath}
\usepackage{float}
\usepackage{multirow}
\usepackage{slashed}
\usepackage{braket}
\usepackage{xcolor}
\usepackage{physics}
\usepackage{multirow}
\usepackage{gensymb}
\usepackage[colorlinks=true, pdfstartview=FitV, bookmarks=true, bookmarksnumbered=true, breaklinks]{hyperref}
\usepackage{mathtools,braket}

\usepackage{lipsum}  
\usepackage{color}
\definecolor{blue}{rgb}{0.0, 0.0, 1.0}
\definecolor{red}{rgb}{1.0, 0.0, 0.0}
\definecolor{royalblue}{rgb}{0.0, 0.14, 0.4}
\hypersetup{linkcolor=blue, citecolor=blue, urlcolor=blue}

\def\orcid#1{\kern .08em\href{https://orcid.org/#1}{\includegraphics[keepaspectratio,width=0.7em]{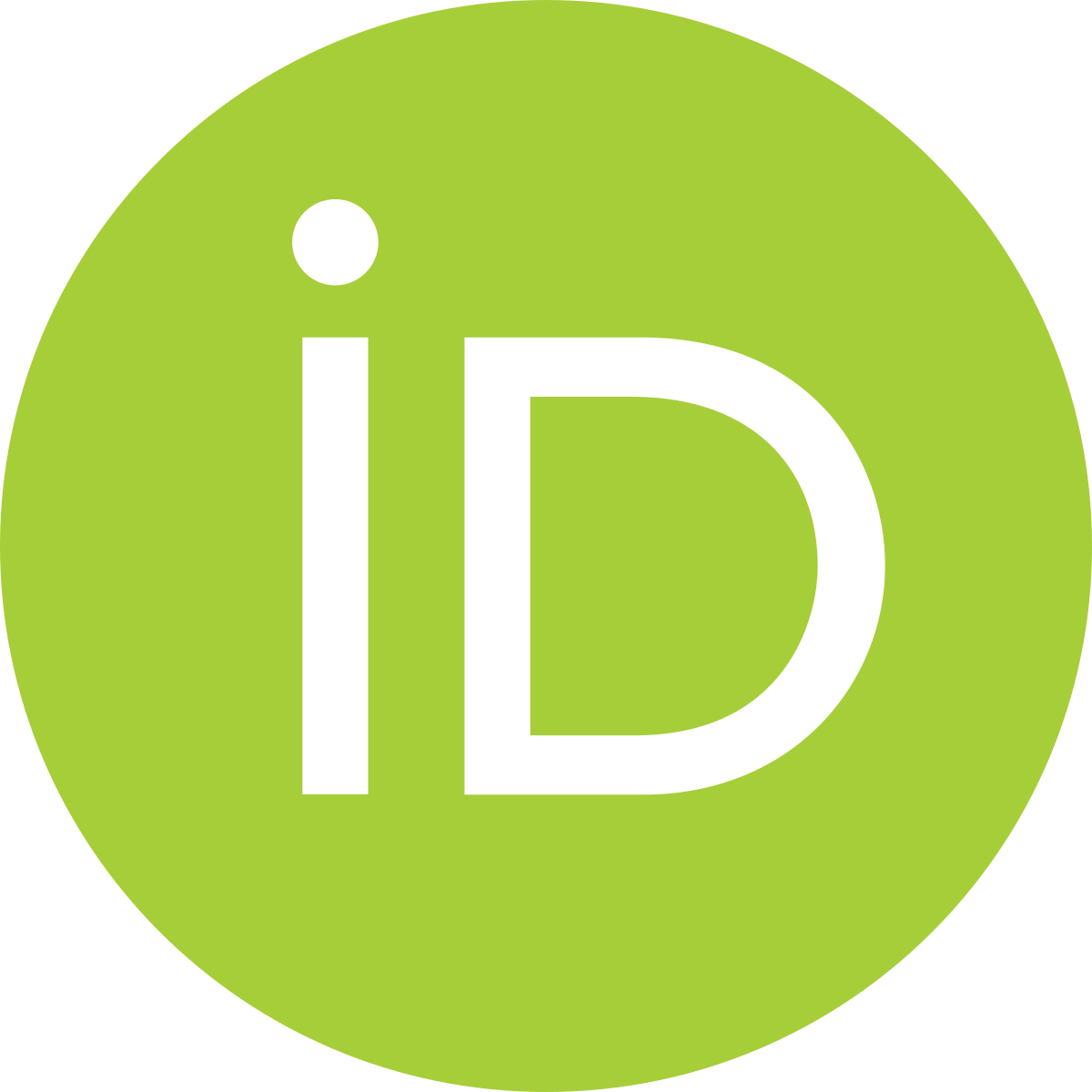}}}

\begin{document}
\title{Two-pion emission decays of negative parity singly heavy baryons}

\author{Nongnapat Ponkhuha\orcid{0009-0004-7891-802X}}
\email{nongnapat.po@kkumail.com}
\affiliation{Department of Physics, Faculty of Science, Khon Kaen University, Khon Kaen 40002, Thailand}

\author{Ahmad Jafar Arifi\orcid{0000-0002-9530-8993}}
\email[Corresponding author: ]{ahmad.arifi@riken.jp}

\affiliation{Few-body Systems in Physics Laboratory, RIKEN Nishina Center, Wako 351-0198, Japan}
\affiliation{Research Center for Nuclear Physics, Osaka University, Ibaraki 567-0047, Japan}

\author{Daris Samart\orcid{0000-0003-4690-7591}}
\email[Corresponding author: ]{darisa@kku.ac.th}
\affiliation{Department of Physics, Faculty of Science, Khon Kaen University, Khon Kaen 40002, Thailand}

\date{\today}

\begin{abstract}
We investigate two-pion emission decays of singly charmed and bottom baryons, focusing on $\Lambda_Q^*(1P)$ and $\Xi_Q^*(1P)$ with $Q=c$ (charm) or $b$ (bottom) quarks and $J^P=1/2^-,3/2^-$, belonging to antisymmetric flavor triplet $\bar{\boldsymbol{3}}_F$. 
Our analysis encompasses both sequential processes, involving intermediate states belonging to symmetric flavor sextet $\boldsymbol{6}_F$ such as $\Sigma_Q(1S)$ and $\Xi_Q^\prime(1S)$ respectively with $J^P=1/2^+,3/2^+$, derived from the chiral quark model, and direct process crucial for comparison with experimental data, whose coupling constants estimated using the chiral-partner scheme. 
We also incorporate the convolution of the parent particle's mass for the Dalitz plot, enabling a more realistic comparison with experimental data. 
We scrutinize the Dalitz plots of these negative parity states in light of recent Belle measurements for $\Lambda_c(2625)^+$. 
Our findings support the assignment of $\Lambda_c(2625)^+$ as the $\lambda$-mode excitation with $J^P=3/2^-$ in the quark model, deduced from the the ${\Lambda_c\pi}$ invariant mass distribution, and we then give predictions for other cases, including the $\Xi_Q^*$ decays. 
The observed asymmetry in the ${\pi\pi}$ invariant mass distribution underscores the important role of the direct process, reflecting the chiral-partner structure in the heavy baryon sector. 
It is evident that the presence of the direct process is not significant in the three-body decays unless the $S$-wave resonance contribution is suppressed. 
We suggest further experimental verification to test our predictions and get more insights into the structure of heavy baryons.
\end{abstract}

\maketitle

%--------------------------------------------------------
\section{Introduction}
%--------------------------------------------------------

\begin{figure}[t]
\centering
\includegraphics[width=0.5\columnwidth]{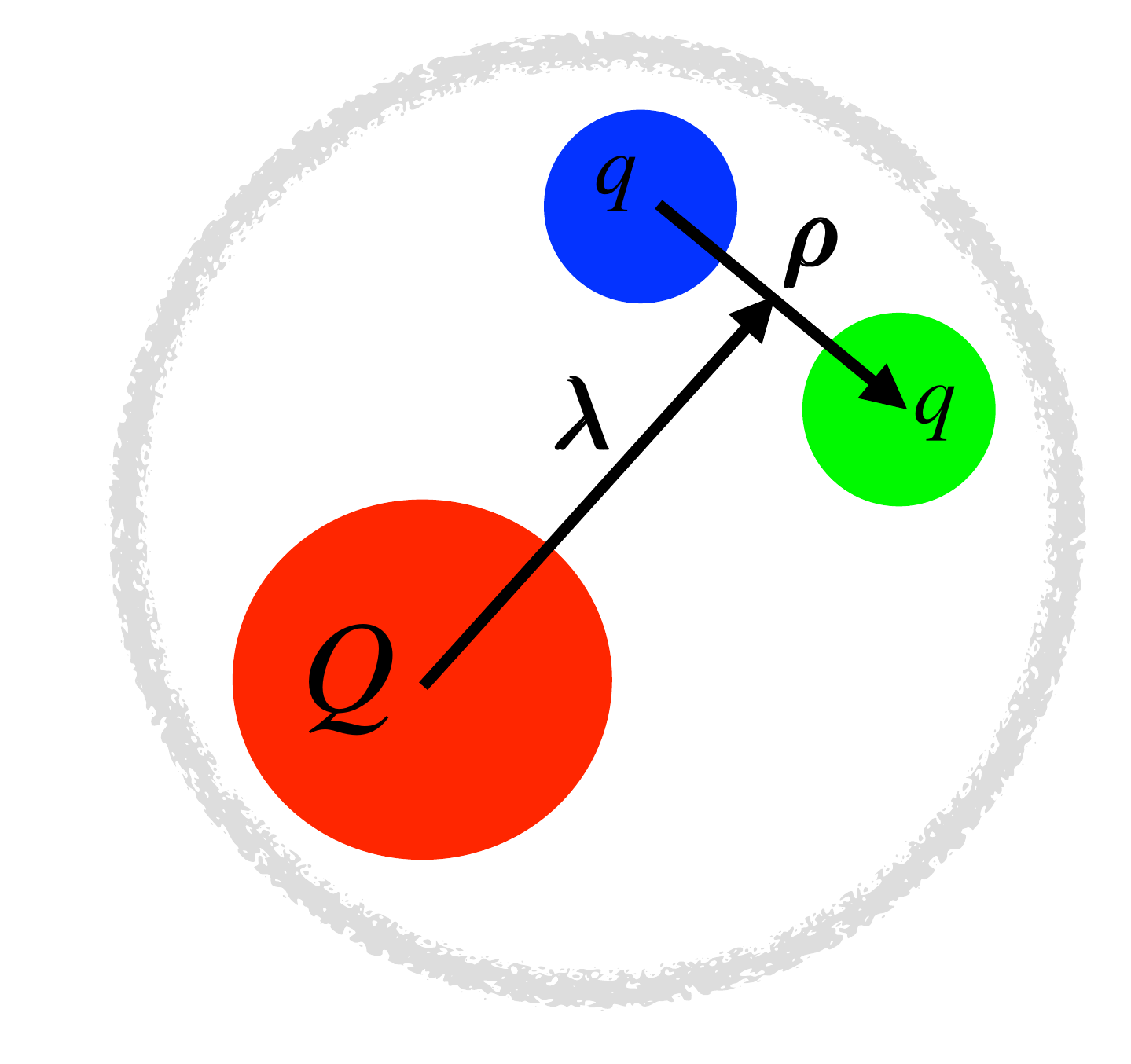}
\caption{\label{fig:baryon} 
Jacobi coordinates $\bm{\lambda}$ and $\bm{\rho}$ of three constituent quarks $(Qqq)$ within a singly heavy baryon. }
\end{figure}

Heavy baryons provide an important platform for studying the interplay between light-quark and heavy-quark dynamics within the framework of quantum chromodynamics (QCD). 
While the dynamics of light quarks are governed by chiral symmetry, in the heavy-quark limit, the spin of heavy quarks decouples from these dynamics, which are then governed by heavy-quark symmetry (HQS)~\cite{Isgur:1991wq}.
This interplay of symmetries plays a crucial role in understanding the structure of heavy baryons.
Compared to light baryons, heavy baryons exhibit a separation between orbital motions, such as $\lambda$ mode $(l_\lambda\neq 0)$ and $\rho$ mode $(l_\rho\neq 0)$, as shown in Fig.~\ref{fig:baryon}, and have relatively smaller widths that facilitate their analysis~\cite{ParticleDataGroup:2022pth}. 

Since the discovery of the $\Lambda_c$, 
many new states have been observed and added to the Particle Data Group (PDG)~\cite{ParticleDataGroup:2022pth}. Although most ground states of singly heavy baryons have been discovered, spectroscopic properties of excited heavy baryon states, such as quantum numbers, mass, width, and branching fractions, remain poorly understood in some cases. Establishing the spectroscopy of heavy baryons requires combined efforts from theory and experiment. Recent experimental advancements~\cite{LHCb:2012kxf,LHCb:2018vuc,LHCb:2019soc,LHCb:2020iby,LHCb:2020lzx,LHCb:2021ssn,LHCb:2023zpu,LHCb:2020lzx,Belle:2022voy,Belle:2008yxs,Belle:2016lhy,Belle:2020ozq,Belle:2020tom,BESIII:2024udz,CDF:2011zbc,CDF:2013pvu,CMS:2020zzv,CMS:2021rvl} coupled with developments of various theoretical models~\cite{Nagahiro:2016nsx, Kim:2022pyq,Yoshida:2015tia,Garcia-Tecocoatzi:2022zrf,Garcia-Tecocoatzi:2023btk, Yang:2022oog, Tan:2023opd, Wang:2017kfr,Ebert:2011kk,Chen:2016iyi,Bijker:2020tns, Arifi:2020ezz,Arifi:2020yfp,Arifi:2021wdf,Suenaga:2022ajn,Suenaga:2021qri,Roberts:2007ni,Garcilazo:2007eh,Gandhi:2019bju,Kawakami:2018olq,Kawakami:2019hpp,Arifi:2017sac, Arifi:2018yhr,Arifi:2021orx,Nieves:2024dcz,Yang:2022oog,Yang:2019cvw,Nieves:2019nol,Zhong:2007gp,Cheng:2006dk,Cho:1994vg,Hyodo:2013iga,Guo:2016wpy,Lu:2016gev,Du:2022rbf,Lu:2014ina} have significantly contributed to our understanding of heavy baryons. 
Interested readers may refer to recent reviews on this topic~\cite{Cheng:2021qpd,Chen:2022asf}. 

Singly heavy baryons $(Qqq)$ are generally classified based on the internal symmetry of their two light quarks~\cite{Chen:2022asf}. 
They fall into antisymmetric $\bar{\bm{3}}_F$ and symmetric $\bm{6}_F$ flavor representations. 
The light quarks can form antisymmetric ($s=0$) or symmetric ($s=1$) spin configurations, with orbital excitations in the $\lambda$ or $\rho$ mode. The overall configuration must be antisymmetric due to the Pauli principle. 
For orbitally excited $(1P)$ states with the $\lambda$ mode and antisymmetric flavor triplet $\bar{\bm{3}}_F$, they have a unique spin configuration $s=0$, resulting in a brown-muck spin $j^p=1^-$. 
Combined with the heavy quark spin $s_Q$, they form a HQS doublet of $J(j)^-=1/2(1)^-$ and $3/2(1)^-$. 
While the low-lying singly heavy baryons are expected to be dominated by the $\lambda$-mode excitations~\cite{Arifi:2018yhr,Yoshida:2015tia}, the $\rho$ mode excitations are believed to have larger masses and their identification in experiments is yet to be resolved. 
Furthermore, their masses being located near the threshold complicates the situation, as non-standard behavior may arise~\cite{Hyodo:2013iga,Guo:2016wpy,Lu:2016gev,Du:2022rbf,Lu:2014ina}.

While analyzing mass spectra is crucial, examining decay patterns offers additional constraints on the structure of singly heavy baryons.
Specifically, the strong decays of excited states, through two-pion emission processes, are of great importance, as they represent a substantial portion of observed decays for $\Lambda_Q^*$ and $\Xi_Q^*$ states~\cite{ParticleDataGroup:2022pth}. Dalitz plots and invariant mass distributions from three-body decays provide additional insights into their structure. 
For instance, in previous analyses~\cite{Arifi:2020ezz,Arifi:2020yfp,Arifi:2021wdf}, the invariant mass distributions for $\Lambda_c(2765)^+$ and $\Lambda_b(6072)^0$ have been useful for determining their spin and parity. 
Additionally, studies on three-body decays of $\Lambda_c(2595)^+$ and $\Lambda_c(2625)^+$ have provided a way to distinguish between $\lambda$ and $\rho$ mode excitations in the quark model~\cite{Arifi:2017sac, Arifi:2018yhr} and expose the possible signature of direct processes in the chiral-partner structure~\cite{Kawakami:2018olq,Kawakami:2019hpp}.

%--------------------------------------------------------
%--------------------------------------------------------

%--------------------------------------------------------
\begin{figure*}[t]
\centering
\includegraphics[width=2\columnwidth]{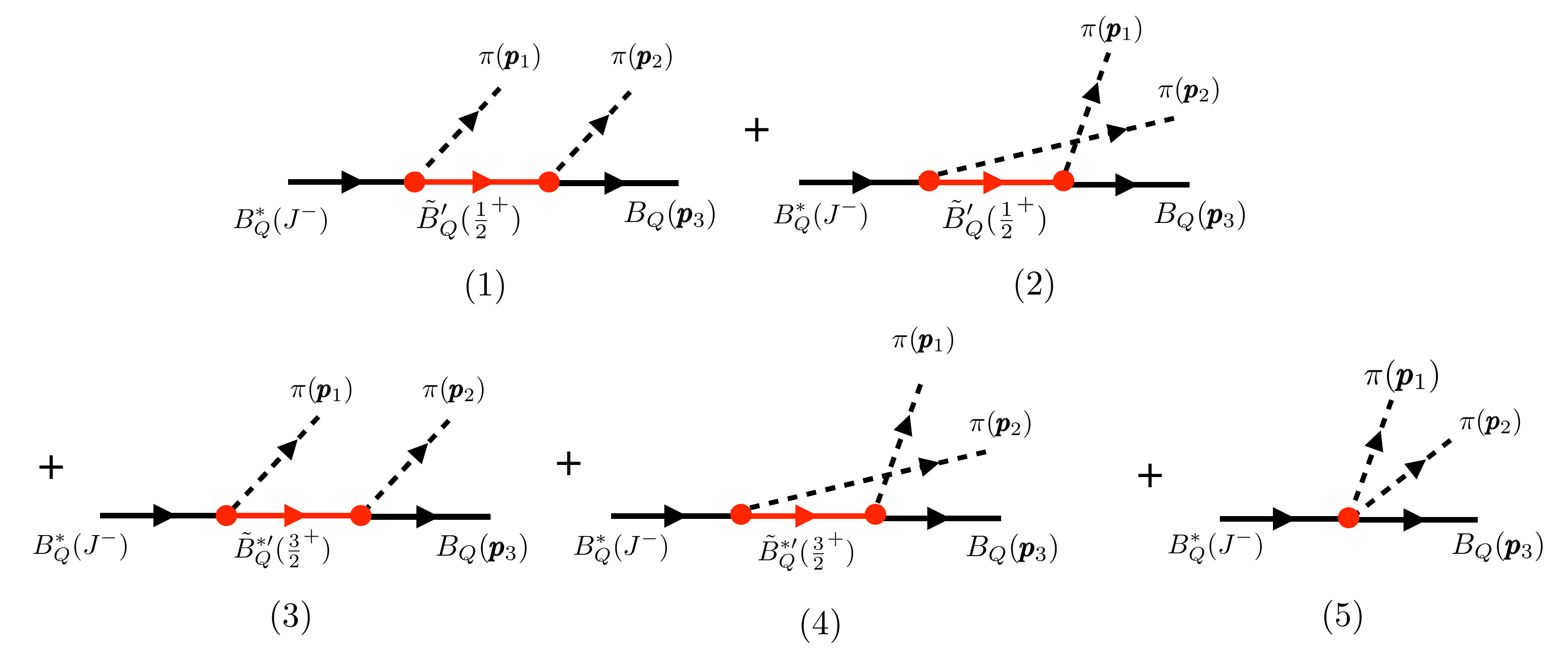}
\caption{\label{fig:diagram} 
Two-pion emission decays of heavy baryons $B_Q^*(J^-)$ where $B_Q=\Lambda_Q$ or $\Xi_Q$ with $Q=c,b$ quark. We consider two cases with $J^P=1/2^-,3/2^-$ of the initial state. The first four diagrams represent the sequential process through intermediate states $\tilde{B}_Q^{(\prime)}(1/2^+)$ and $\tilde{B}_Q^{*(\prime)}(3/2^+)$, while the last diagram represents the direct two-pion coupling.}
\end{figure*}
%--------------------------------------------------------

Recent experimental progress~\cite{CMS:2021rvl,LHCb:2023zpu} has discovered states possibly assigned to the $1P$ state doublets in the SU(3) flavor $\bar{\bm{3}}_F$, prompting systematic studies of these states. Understanding the structure of $\Lambda_Q^*$ and $\Xi_Q^*$ baryons requires a thorough analysis of their two-pion emission decays, especially in light of new data~\cite{Belle:2022voy}.
Motivated by that, we extend previous studies~\cite{Arifi:2017sac, Arifi:2018yhr} to systematically analyze the two-pion emission decays of $\Lambda_Q^*(1P)$ and $\Xi_Q^*(1P)$, aiming to further elucidate their internal structure. We employ the chiral quark model to calculate coupling constants for sequential processes~\cite{Nagahiro:2016nsx}, while estimating the coupling for the direct process using chiral-partner structure~\cite{Kawakami:2018olq,Kawakami:2019hpp}. Our analysis also includes the effect of finite widths of parent particles on the Dalitz plot~\cite{Arifi:2020ezz,Arifi:2020yfp} to enable a more realistic comparison with experimental data.

Our calculated invariant mass distributions are found to be consistent with recent Belle data on $\Lambda_c(2625)^+$~\cite{Belle:2022voy}, further supporting the $\lambda$-mode assignment in the quark model for this state. 
Specifically, the ratio between the peak of $\Sigma_c(2455)$ and the shoulder of $\Sigma_c(2520)$ reflected in the $\Lambda_c\pi$ invariant mass is well reproduced using the $\lambda$ mode, while that with the $\rho$ mode is not suitable with the observed shoulder by the $\Sigma_c(2520)$ resonance~\cite{Arifi:2017sac, Arifi:2018yhr}. 
Moreover, the direct two-pion coupling estimated using the chiral-partner scheme~\cite{Kawakami:2018olq, Kawakami:2019hpp} would result in an asymmetrical structure in the $\pi\pi$ invariant mass distribution as clearly seen in the Belle data~\cite{Belle:2022voy}. 

With the success of explaining the $\Lambda_c(2625)^+$ decay with the $\lambda$ assignment, we provide predictions for other cases, such as $\Lambda_b$ and $\Xi_{c(b)}$ decays.
We suggest that the direct process may not play a significant role unless the $S$-wave resonant contribution is suppressed as in the case of $\Lambda_c(2625)^+$. 
Our predictions for $\Lambda_b$ and $\Xi_{c(b)}$ decays can be tested by experiments such as LHCb, BelleII, and BESIII, shedding light on the structure of heavy baryons.

The rest of the paper is structured as follows. 
In Section~\ref{sec:model}, we present the model of three-body decay and outline how we determine coupling constants from the chiral quark model and chiral-partner structure. 
Section~\ref{sec:result} is dedicated to numerical results and discussion, 
followed by our conclusions in Section~\ref{sec:conclusion}.

%--------------------------------------------------------
\section{Model Description}
%--------------------------------------------------------
\label{sec:model}

In this section, we will explain the computation of two-pion emission decay amplitudes of anti-symmetric flavor triplet $\bar{\bm{3}}_F$ heavy baryons $B_Q^*(1P)\rightarrow B_Q \pi\pi$ through the intermediate symmetric flavor sextet $\bm{6}_F$ heavy baryons $\tilde{B}_Q(1S)$ and the direct process as illustrated in Fig.~\ref{fig:diagram}. 
We then show how we determine the coupling constant for each vertex and compute the Dalitz plot.

%--------------------------------------------------------
\subsection{Three-body decay amplitudes}
%--------------------------------------------------------

In general, we consider five Feynman diagrams including four sequential process and one direct process, as shown in Fig.~\ref{fig:diagram}, unless the decay channel does not exist, for instance the cross diagram is not possible for some $\Xi_Q$ decays due to some the charge conservation. 
For the $B_Q^*(1/2^-)\to B_Q\pi\pi$, the decay amplitudes are written as~\cite{ Arifi:2018yhr} 
\begin{eqnarray}
    \mathcal{M}_1(\tilde{B}_Q) &=& \frac{g_{11}^a g^b_{11} N \chi_{B_Q}^\dagger (\bm{\sigma} \cdot \bm{p}_2) \chi_{B_Q^*}}{m_{\pi_2B_Q} - M_{\tilde{B}_Q(\frac{1}{2}^+)} + \frac{i}{2}\Gamma_{\tilde{B}_Q(\frac{1}{2}^+)}}, \label{eq:first}\\ 
    \mathcal{M}_2(\tilde{B}_Q^*) &=& \frac{g_{13}^a g^b_{31} N \chi_{B_Q}^\dagger (\bm{S} \cdot \bm{p}_2)(\bm{S}^\dagger \cdot \bm{p}_1) (\bm{\sigma} \cdot \bm{p}_1) \chi_{B_Q^*}}{m_{\pi_2B_Q} - M_{\tilde{B}_Q(\frac{3}{2}^+)} + \frac{i}{2}\Gamma_{\tilde{B}_Q(\frac{3}{2}^+)}}, \\ 
    \mathcal{M}_3(\tilde{B}_Q^{\prime}) &=& \frac{g_{11}^{a\prime} g_{11}^{b\prime} N \chi_{B_Q}^\dagger (\bm{\sigma} \cdot \bm{p}_1) \chi_{B_Q^*}}{m_{\pi_1 B_Q} - M_{\tilde{B}_Q^\prime(\frac{1}{2}^+)} + \frac{i}{2}\Gamma_{\tilde{B}_Q^\prime(\frac{1}{2}^+)}}, \\ 
    \mathcal{M}_4(\tilde{B}_Q^{*\prime}) &=& \frac{g_{13}^{a\prime} g_{31}^{b\prime} N \chi_{B_Q}^\dagger (\bm{S} \cdot \bm{p}_1)(\bm{S}^\dagger \cdot \bm{p}_2)(\bm{\sigma} \cdot \bm{p}_2) \chi_{B_Q^*}}{m_{\pi_1B_Q} - M_{\tilde{B}_Q^{*\prime}(\frac{3}{2}^+)} + \frac{i}{2}\Gamma_{\tilde{B}_Q^{*\prime}(\frac{3}{2}^+)}}, \quad\quad \\ 
    \mathcal{M}_5(\mathrm{dir}) &=& \frac{g_{11}^d N }{f_\pi} \chi_{B_Q}^\dagger (\bm{\sigma} \cdot (\bm{p}_1 + \bm{p}_2)) \chi_{B_Q^*}.
\end{eqnarray}
For the $B_Q^*(3/2^-)\to B_Q\pi\pi$, the decay amplitudes are written as~\cite{ Arifi:2018yhr} 
\begin{eqnarray}
    \mathcal{M}_1(\tilde{B}_Q) &=& \frac{g_{31}^a g^b_{11} N\chi_{B_Q}^\dagger (\bm{\sigma} \cdot \bm{p}_2) (\bm{\sigma} \cdot \bm{p}_1) (\bm{S} \cdot \bm{p}_1) \chi_{B_Q^*}}{m_{\pi_2B_Q} - M_{B_Q(\frac{1}{2}^+)} + \frac{i}{2}\Gamma_{B_Q(\frac{1}{2}^+)}}, \\ 
    \mathcal{M}_2(\tilde{B}_Q^*) &=& \frac{g_{33}^a g^b_{31} N\chi_{B_Q}^\dagger (\bm{S} \cdot \bm{p}_2) \chi_{B_Q^*}}{m_{\pi_2B_Q} - M_{B_Q^*} + \frac{i}{2}\Gamma_{B_Q^*(\frac{3}{2}^+)}}, \\ 
    \mathcal{M}_3(\tilde{B}_Q^\prime) &=& \frac{g_{31}^{a\prime} g_{11}^{b\prime} N\chi_{B_Q}^\dagger (\bm{\sigma} \cdot \bm{p}_1) (\bm{\sigma} \cdot \bm{p}_2) (\bm{S} \cdot \bm{p}_2) \chi_{B_Q^*}}{m_{\pi_1 B_Q} - M_{B_Q^\prime(\frac{1}{2}^+)} + \frac{i}{2}\Gamma_{B_Q^\prime(\frac{1}{2}^+)}}, \quad\quad \\ 
    \mathcal{M}_4(\tilde{B}_Q^{*\prime}) &=& \frac{g_{33}^{a\prime} g_{31}^{b\prime} N\chi_{B_Q}^\dagger (\bm{S} \cdot \bm{p}_1)\chi_{B_Q^*}}{m_{\pi_1B_Q} - M_{B_Q^{*\prime}(\frac{3}{2}^+)} + \frac{i}{2}\Gamma_{B_Q^{*\prime}(\frac{3}{2}^+)}}, \\ 
    \mathcal{M}_5(\mathrm{dir}) &=& \frac{g_{31}^d N}{f_\pi}\chi_{B_Q}^\dagger (\bm{S} \cdot (\bm{p}_1 + \bm{p}_2)) \chi_{B_Q^*}, \label{eq:second}
\end{eqnarray}
with $N= \sqrt{2M_{B_Q^*} } \sqrt{2M_{B_Q}}$, and $\chi_{B_Q^*}$ and $\chi_{B_Q}^\dagger$ the spin states of the initial and final state of heavy baryon. 
The $\tilde{B}_Q^{(\prime)}$ and $\tilde{B}_Q^{*(\prime)}$ are the intermediate states, belonging to the symmetric flavor sextet $\bm{6}_F$ of heavy baryon, and their mass distribution is modeled by a simple Breit-Wigner function. The primed symbol corresponds to the baryon with different charge according to the decay process.
The mass and width parameters of the heavy baryons considered here are given in Table~\ref{tab:mass}.

The notation of the particle number follows that in Fig.~\ref{fig:diagram}, where $\bm{p}_1$, $\bm{p}_2$, and $\bm{p}_3$ are defined for each final state. 
The $\bm{\sigma}$ and $\bm{S}$ are the usual Pauli matrix and the spin transition matrix from spin 3/2 to 1/2.
We also define the invariant mass of two systems as $m_{\pi_1 B_Q}$ and $m_{\pi_2 B_Q}$ where $\pi_1(\bm{p}_1)$ and $\pi_2(\bm{p}_2)$ are the first and outgoing pions, respectively.
The other invariant mass of $m_{\pi_1\pi_2}$ is constrained through a relation 
\begin{eqnarray}
    m_{\pi_1\pi_2}^2 + m_{\pi_1B_Q}^2 + m_{\pi_2B_Q}^2 = M_{B_Q^*}^2 + M_{B_Q}^2 + M_{\pi_1}^2 + M_{\pi_2}^2.\nonumber\\
\end{eqnarray}
We note that $\bm{p}_1$ and $\bm{p}_2$ are functions of the invariant mass and are evaluated in the rest frame of the initial $B_Q^*$ state
\begin{eqnarray}
    |\bm{p}_1|^2 &=& \frac{\lambda(M_{B_Q^*}, M_{\pi_1},m_{\pi_2B_Q})}{4M^2_{B_Q^*}}, \\
    |\bm{p}_2|^2 &=& \frac{\lambda(M_{B_Q^*}, M_{\pi_2},m_{\pi_1B_Q})}{4M^2_{B_Q^*}},
\end{eqnarray}
where $\lambda(x,y,z)=x^2 + y^2 + z^2 -2(xy + xz + yz)$. 

In Eqs.~\eqref{eq:first}-\eqref{eq:second}, the coupling constants are written as $g_{2J,2J^\prime}$, showing the total angular momentum of the initial $J$ and final $J^\prime$ state for each vertex in the subscript.
The superscript labels $a$ and $b$ indicate the first and second vertex, while label $d$ represents the direct process. 
Also, the primed coupling constants indicate the crossed diagram processes in Fig.~\ref{fig:diagram}~(2) and (4).
Furthermore, the direct two-pion amplitude includes the pion decay constant $f_\pi=93$ MeV to match the order of chiral expansion.
In the next subsection, we show that the coupling strength extracted from the quark model contain the Gaussian form factor for each vertex which simulate the finite size of the heavy baryon.

In general, due to the $S$-wave nature decay of the first vertex, it is clearly seen that $\mathcal{M}_1(\tilde{B}_Q)$ and  $\mathcal{M}_3(\tilde{B}_Q^\prime)$ are expected to be the dominant processes for $B_Q^*(1/2^-)\to B_Q\pi\pi$ while $\mathcal{M}_2(\tilde{B}_Q^*)$ and  $\mathcal{M}_4(\tilde{B}_Q^{*\prime})$ are suppressed due to the $D$-wave decay.
On the other hand, $\mathcal{M}_2(\tilde{B}_Q^*)$ and  $\mathcal{M}_4(\tilde{B}_Q^{*\prime})$ are expected to be dominant in the $B_Q^*(3/2^-)\to B_Q\pi\pi$ due to $S$-wave decay.
However, their actual contribution depends on the process and kinematics that will be discussed in detail later.

The total amplitude is a coherent sum of the amplitudes given by
\begin{eqnarray}
    \mathcal{M}_\mathrm{total} &=& \sum_{i=1}^5 \mathcal{M}_i,
\end{eqnarray}
where relative phase is absorbed in the coupling constant determined from the quark model~\cite{Nagahiro:2016nsx}.
The three-body decay width is then calculated as~\cite{ParticleDataGroup:2022pth}
\begin{eqnarray}
\Gamma &=& \frac{(2\pi)^4}{2M_{B_Q^*}} \int  \overline{|\mathcal{M}_\mathrm{total}|^2}\ \dd \Phi_3(P_{B_Q^*};p_{1},p_{2},p_{3})\nonumber\\
    &=& \frac{1}{32M^3_{B_Q^*}(2\pi)^3} \int \overline{|\mathcal{M}_\mathrm{total}|^2}\ \dd m^2_{\pi_1\pi_2} \dd m^2_{\pi_2B_Q},\quad \label{eq:dalitz}
\end{eqnarray}
where ${\rm d}\Phi_3$ is the three-body phase space.
The three-body decay can be described by a two-dimensional plot of invariant masses $m^2_{\pi_1\pi_2}$ and $m^2_{\pi_2B_Q}$.
By integrating one of those invariant masses, we can obtain the invariant mass distribution.
Note that we should divide the squared amplitude by a symmetric factor when there are two identical final states such as in the $B_Q \pi^0\pi^0$ channel.

The Dalitz boundary of three-body decays can be computed as
\begin{eqnarray}
 (m_{\pi_1\pi_2})_\pm^2 &=& M_{\pi_1}^2 + M_{\pi_2}^2 + 2\left(E_{1}^* E_{2}^* \pm |\bm{p}_{1}^*| |\bm{p}_{2}^*| \right), \quad \quad
\end{eqnarray}
where the energy $E_i^*$ and momentum $\bm{p}_{i}^*$ are evaluated in the rest frame of $\pi_2$ and $B_Q$.
The plot of $(m_{\pi_1\pi_2})_+^2$ (upper boundary) and $(m_{\pi_1\pi_2})_-^2$ (lower boundary) as a function of $m_{\pi_2 B_Q}^2$ will produce the total Dalitz boundary, inside of which the three-body decay is allowed.
The minimal and maximum points of the Dalitz plot in entire region of $m_{\pi_2 B_Q}^2$ are given by
\begin{eqnarray}
    (m_{\pi_1\pi_2})_\mathrm{max}^2 &=& (M_{B_Q^*} - M_{B_Q})^2,\\
    (m_{\pi_1\pi_2})_\mathrm{min}^2 &=& (M_{\pi_1} + M_{\pi_2})^2.
\end{eqnarray}

In previous works~\cite{Arifi:2017sac, Arifi:2018yhr}, all of the Dalitz plots and other observables are obtained by choosing a fixed value of the initial mass.
Here, we present the results by taking into account the effect of finite width of the initial state which would smear the Dalitz plot, providing a more realistic comparison with experimental data.
Although the initial states $B_Q^*(1P)$ have relatively narrow width as compared with, \textit{e.g.}, $\Lambda_c^*(2765)$, the effects are still not negligible.
By doing this, the Dalitz boundary becomes larger as shown in Fig.~5 of Ref.~\cite{Arifi:2020ezz} and smear the resonance bands.

To perform the above procedure, we use a Breit-Wigner form to model the mass distribution of the initial $B_Q^*$; 
\begin{eqnarray}
\tilde{\Gamma}  = \frac{1}{\tilde{N}} \int_{M_{B_Q^*}-2\Gamma_{B_Q^*}}^{M_{B_Q^*}+2\Gamma_{B_Q^*}}  \frac{\Gamma(m_{B_Q^*})\ \dd m_{B_Q^*}}{(m_{B_Q^*} - M_{B_Q^*} )^2 + \Gamma_{B_Q^*}^2/4}, \quad
\end{eqnarray}
where $\Gamma(m_{B_Q^*})$ is the calculated decay width of $B_Q^*$ which depends on the variable mass $m_{B_Q^*}$.
The normalization factor $\tilde{N}$ is defined by
\begin{eqnarray}
\tilde{N} =  \int_{M_{B_Q^*}-2\Gamma_{B_Q^*}}^{M_{B_Q^*}+2\Gamma_{B_Q^*}} \frac{\dd m_{B_Q^*}}{(m_{B_Q^*} - M_{B_Q^*} )^2 + \Gamma_{B_Q^*}^2/4}.
\end{eqnarray}
We have used PDG~\cite{ParticleDataGroup:2022pth} and recently observed value for the mass ($M_{B_Q^*}$) and width ($\Gamma_{B_Q^*}$) of $B_Q^*$ as presented in Table~\ref{tab:mass}.
In the calculation, there are some states that are not yet observed in experiments.
In that case, we assume the mass to its isospin partner for simplicity.

%--------------------------------------------------------
\begin{table}[t]
\centering
\caption{ Mass $(M)$, width $(\Gamma)$, and spin-parity $(J^P)$ of the initial, intermediate, and final states of heavy baryons considered in the present work. The values of $M$ and $\Gamma$ are taken from PDG~\cite{ParticleDataGroup:2022pth}. The pion mass is also taken from PDG.}
\label{tab:mass}
\renewcommand{\arraystretch}{1.15}
\begin{ruledtabular}
\begin{tabular}{lccc}
States              & $J^P$  & $M$ [MeV] & $\Gamma$ [MeV] \\ \hline
\multicolumn{4}{l}{$B_Q^*$ (Initial state, $\bar{\bm{3}}_F$)} \\ 
$\Lambda_c(2595)^+$ & $1/2^-$ & $2592.25\pm 0.28$ & $2.6 \pm 0.6$ \\
$\Lambda_c(2625)^+$ & $3/2^-$ & $2628.11\pm 0.19$ & $<0.97$ \\ 
$\Lambda_b(5912)^0$ & $1/2^-$ & $5912.19\pm 0.17$ & $<0.25$ \\
$\Lambda_b(5920)^0$ & $3/2^-$ & $5920.09\pm 0.17$ & $<0.19$ \\ 
$\Xi_c(2790)^+$     & $1/2^-$ & $2791.9\pm 0.5$ & $8.9\pm 1.0$ \\
$\Xi_c(2790)^0$     & $1/2^-$ & $2793.9\pm 0.5$ & $10.0\pm 1.1$  \\
$\Xi_c(2815)^+$     & $3/2^-$ & $2816.51\pm 0.25$ & $2.43\pm 0.26$ \\ 
$\Xi_c(2815)^0$     & $3/2^-$ & $2819.79\pm 0.30$ & $2.54\pm 0.25$ \\ 
$\Xi_b(6087)^0$     & $1/2^-$ & $6087.24\pm0.20$ & $2.43 \pm 0.51$  \\
$\Xi_b(1/2^-)^-$    & \multicolumn{3}{l}{Not yet observed} \\
$\Xi_b(6095)^0$     & $3/2^-$ & $6095.36\pm 0.15$ & $0.50\pm 0.33$ \\ 
$\Xi_b(6100)^-$     & $3/2^-$ & $6099.74\pm 0.11$ & $0.94\pm 0.30$\\  \hline
\multicolumn{4}{l}{$\tilde{B}_Q^{(*)}$ (Intermediate state, $\bm{6}_F$)} \\ 
$\Sigma_c(2455)^{++}$  & $1/2^+$ & $2453.97\pm 0.14$ & $1.89^{+0.09}_{-0.18}$ \\
$\Sigma_c(2455)^{+}$  & $1/2^+$ & $2452.65^{+0.22}_{-0.16}$ & $2.3\pm 0.4$\\
$\Sigma_c(2455)^{0}$  & $1/2^+$ & $2453.75\pm 0.14$ & $1.83^{+0.11}_{-0.19}$ \\
$\Sigma_c(2520)^{++}$  & $3/2^+$ & $2518.41^{+0.22}_{-0.18}$ & $14.78^{+0.30}_{-0.40}$ \\
$\Sigma_c(2520)^{+}$  & $3/2^+$ & $2517.4^{+0.7}_{-0.5}$ & $17.2^{+4.0}_{-2.2}$ \\
$\Sigma_c(2520)^{0}$  & $3/2^+$ & $2518.48\pm 0.20$ & $15.3^{+0.4}_{-0.5}$ \\ 
$\Sigma_b^{+}$  & $1/2^+$ & $5810.56\pm 0.25$ & $5.0\pm 0.5$ \\
$\Sigma_b(1/2^+)^{0}$  & \multicolumn{3}{l}{Not yet observed} \\
$\Sigma_b^{-}$  & $1/2^+$ & $5815.64\pm 0.27$ & $5.3\pm 0.5$  \\
$\Sigma_b^{*+}$  & $3/2^+$ & $5830.32\pm 0.27$ & $9.4\pm 0.5$ \\
$\Sigma_b^*(3/2^+)^{0}$ & \multicolumn{3}{l}{Not yet observed} \\
$\Sigma_b^{*-}$  & $3/2^+$ & $5834.74\pm 0.30$ & $10.4 \pm 0.8$ \\ 
$\Xi_c^{\prime +}$  & $1/2^+$ & $2578.2\pm 0.5$ & \dots \\
$\Xi_c^{\prime 0}$  & $1/2^+$ & $2578.7\pm 0.5$ & \dots \\
$\Xi_c(2645)^{+}$  & $3/2^+$ & $2645.10\pm 0.30$ & $2.14\pm 0.19$  \\
$\Xi_c(2645)^{0}$  & $3/2^+$ & $2646.16\pm 0.25$ & $2.35\pm 0.22$ \\
$\Xi_b^\prime(1/2^+)^{0}$ & \multicolumn{3}{l}{Not yet observed} \\
$\Xi_b^\prime(5935)^{-}$  & $1/2^+$ &  $5935.02\pm0.05$ & $<0.08$ \\
$\Xi_b(5945)^{0}$  & $3/2^+$ & $5952.3\pm0.6$ & $0.90\pm 0.18$ \\
$\Xi_b(5955)^{-}$  & $3/2^+$ & $5955.33\pm 0.13$ & $1.65\pm 0.33$ \\ \hline
\multicolumn{4}{l}{$B_Q$ (final state, $\bar{\bm{3}}_F$)} \\ 
$\Lambda_c^+$ & $1/2^+$ & $2286.46\pm 0.14$ & \dots \\ 
$\Lambda_b^0$ & $1/2^+$ & $5619.60\pm 0.17$ & \dots  \\ 
$\Xi_c^+$     & $1/2^+$ & $2467.71\pm0.23$ & \dots \\ 
$\Xi_c^0$     & $1/2^+$ & $2470.44\pm 0.28$ & \dots  \\ 
$\Xi_b^-$     & $1/2^+$ & $5797.0\pm 0.6$ & \dots  \\
$\Xi_b^0$   & $1/2^+$ &  $5791.9\pm 0.5$ & \dots \\ 
\end{tabular}
\renewcommand{\arraystretch}{1.0}
\end{ruledtabular}
\end{table}
%--------------------------------------------------------

%--------------------------------------------------------
\begin{figure}[t]
\centering
\includegraphics[width=0.9\columnwidth]{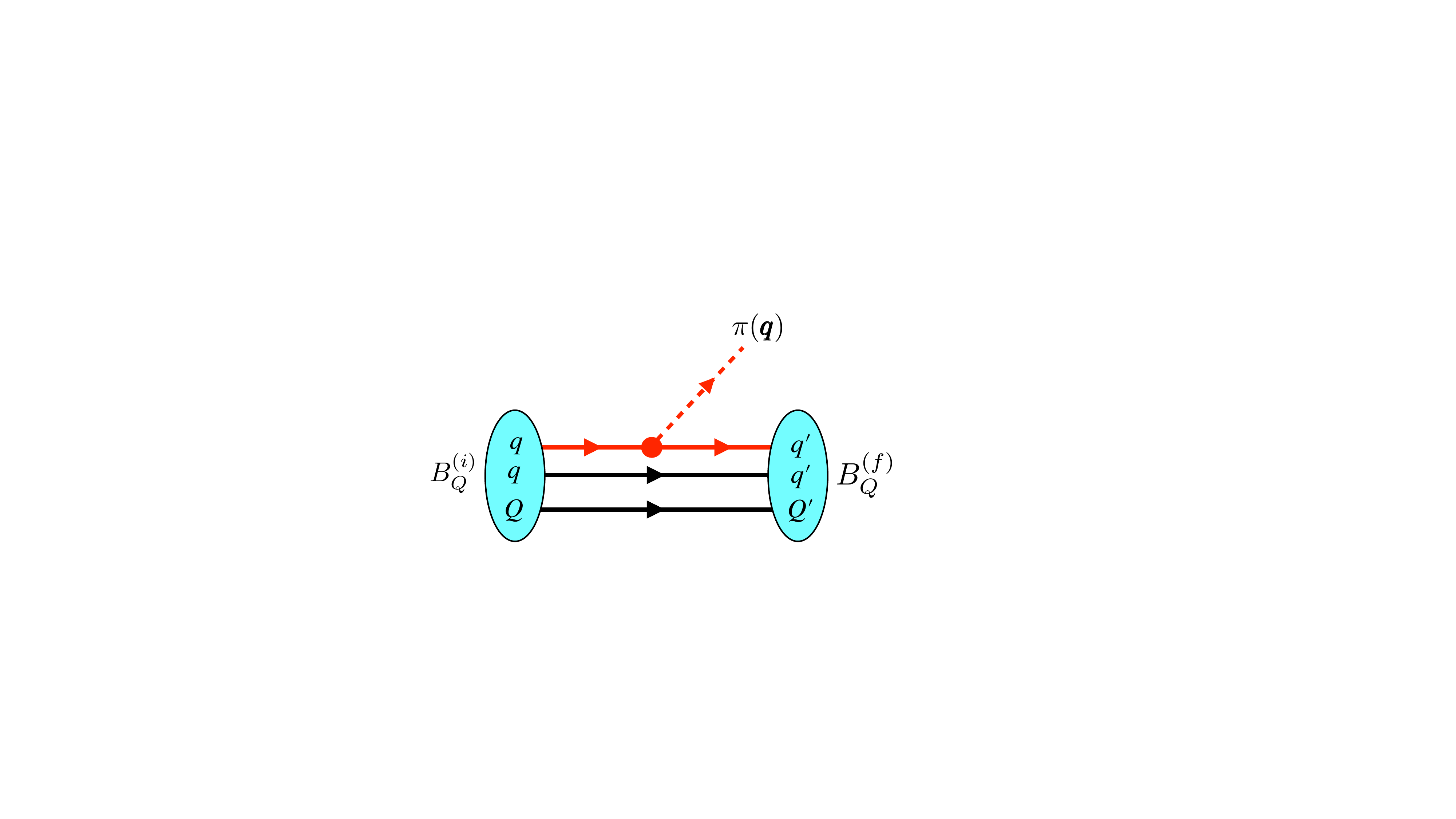}
\caption{\label{fig:diagram2} 
One-pion emission decay of singly heavy baryons in the chiral quark model~\cite{Nagahiro:2016nsx}. The outgoing pion couples to the light quarks, while the heavy quark acts as a spectator.}
\end{figure}
%--------------------------------------------------------

%--------------------------------------------------------
\subsection{Determination of coupling constants}
%--------------------------------------------------------

While the coupling constants of the sequential process are extracted from the chiral quark model~\cite{Arifi:2017sac, Arifi:2018yhr}, the coupling constants for the direct two-pion coupling are derived from the chiral-partner scheme~\cite{Kawakami:2018olq, Kawakami:2019hpp}. 
We will see later that this procedure yields a reasonable description of the experimental data.

%--------------------------------------------------------
\subsubsection{Chiral quark model}
%--------------------------------------------------------

In the chiral quark model~\cite{Nagahiro:2016nsx}, the one-pion emission decay is described in Fig.~\ref{fig:diagram2}. We employ the axial-vector type coupling for the chiral interaction between the outgoing pion and a light quark with mass $m_q$ inside the singly heavy baryon, as given by~\cite{Nagahiro:2016nsx}
\begin{eqnarray}
\mathcal{L}_{\pi qq} = \frac{g_A^q}{2f_\pi} \bar{q} \gamma_\mu\gamma_5 \bm{\tau} q \cdot \partial^\mu \bm{\pi} \label{inter}
\end{eqnarray}
where $g^q_A=1$ is the quark axial vector coupling constant. The heavy quark with mass $m_Q$ acts as a spectator during the decay process. 
Note that we need to account for the case where the outgoing pion couples to the second light quark.

We then perform the nonrelativistic reduction and obtain the one-pion emission decay operator as~\cite{Nagahiro:2016nsx}
\begin{eqnarray}
    \mathcal{H}_{\pi qq} = \frac{g_A^q}{2f_\pi} \left[ \bm{\sigma}\cdot \bm{q} + \frac{\omega}{2m_q}(\bm{\sigma}\cdot \bm{q} - 2\bm{\sigma}\cdot \bm{p}_i)\right]
\end{eqnarray}
where the outgoing pion energy is given by ($\omega,\bm{q}$). Here, $\bm{p}_i$ denotes the initial momentum of the struck light quark. One may also include the next leading order term that would result in a strong suppression to the ground state decay, but the effect is minimal for the negative parity state~\cite{Arifi:2021orx, Arifi:2022ntc}. 
Decay amplitudes, which are expressed in helicity amplitudes $A_h$, are then computed by sandwiching $\mathcal{H}_{\pi qq}$ between the initial and final states of heavy baryon wave functions obtained with the harmonic oscillator (HO) confinement for simplicity~\cite{Nagahiro:2016nsx}.
It is worth noting that a more realistic Hamiltonian and wave function can be obtained by solving the Schrödinger equation using the Gaussian-expansion method (GEM)~\cite{Yoshida:2015tia}. While this approach improves the asymptotic behavior of the wave function~\cite{Arifi:2024mff}, the impact may not be substantial if the size of the HO wave function is properly adjusted.

For the $P$-wave decay of $\tilde{B}_Q^{(*)} \to B_Q \pi$, 
the helicity amplitude $A_h$ is given by~\cite{Nagahiro:2016nsx}
\begin{eqnarray}
-i A_{1/2} = q \mathcal{O}_1 c_1 \tau,
\end{eqnarray}
where $q=|\bm{q}|$ and the factor $\mathcal{O}_1$ is defined by
\begin{eqnarray}
\mathcal{O}_1 = G_a\left( 2 + \frac{\omega}{2m_q+m_Q}\right) F(q^2),
\end{eqnarray}
with a factor
\begin{eqnarray}
G_a = \frac{g_A^q}{2f_\pi} \sqrt{2 M_{B_Q}} \sqrt{2M_{\tilde{B}_Q^{(*)}}},
\end{eqnarray}
and a Gaussian form factor $F(q_\pi^2)$ given by
\begin{eqnarray}
F(q^2) = \mathrm{e}^{-q^2_\lambda/4a^2_\lambda} \mathrm{e}^{-q^2_\rho/4a^2_\rho},
\end{eqnarray}
with $a_\lambda$ and $a_\rho$ being the inverse of the range parameters of HO wave function. 
The momentum transfers for the \(\lambda\) and \(\rho\) modes are given by
\begin{eqnarray}
q_\lambda = q \frac{ m_Q}{2m_q+m_Q},\qquad q_\rho = \frac{q}{2}.
\end{eqnarray}
The coefficient $c_1$ reads
\begin{eqnarray}
c_1 =  \begin{cases}
-1/\sqrt{3} & \text{for}~\tilde{B}_Q(1/2^+), \\
\sqrt{2/3}  & \text{for}~\tilde{B}_Q^{*}(3/2^+).
\end{cases}
\end{eqnarray}
The isospin factor $\tau$ is given by
\begin{eqnarray}
\tau = 
\begin{cases}
1/2 & \text{for}~\Xi_Q\pi^0, \\
1/\sqrt{2} & \text{for}~\Xi_Q \pi^\pm,
\end{cases}
\end{eqnarray}
while $\tau$ is unity for $\Sigma_Q \to \Lambda_Q \pi$ decays. Due to the isospin factor, the decay width (square of the amplitude) of the charged pion emission decay of $\Xi_Q^\prime$ is roughly twice as large as that of the neutral one.

In this work, we consider the $\lambda$-mode excitation as the $\rho$-mode one has been discussed in previous study~\cite{Arifi:2017sac,Arifi:2018yhr,Yoshida:2015tia}, 
but the $\lambda$ mode is preferred for describing the mass and total decay width of low-lying excited states.
We will see later that the $\lambda$-mode can provide a good description of the three-body decays.
The helicity amplitude of $B_Q^{*} (J^-) \to \tilde{B}_Q^{(*)}\pi$ is given by~\cite{Nagahiro:2016nsx}
\begin{eqnarray}
-i A_h = (\mathcal{O}_0^\lambda c_0 + q^2 \mathcal{O}_2^\lambda c_2) \tau.
\end{eqnarray}
We define some factors as
\begin{eqnarray}
\mathcal{O}_0^\lambda &=& \frac{iG_b a_\lambda \omega }{m_q} F(q^2),\\
\mathcal{O}_2^\lambda &=& \frac{iG_b m_Q}{a_\lambda (2m_q+m_Q)}\left( 2 + \frac{\omega}{2m_q+m_Q} \right) F(q^2),
\end{eqnarray}
where
\begin{eqnarray}
    G_b = \frac{g_A^q}{2f_\pi} \sqrt{2 M_{B_Q^*}} \sqrt{2M_{\tilde{B}_Q^{(*)}}}.
\end{eqnarray}
The coefficients $c_0$ and $c_2$ for the $B_Q^*(1/2^-)$ decay with $h=1/2$ are given by
\begin{eqnarray}
    c_0 = -\frac{1}{\sqrt{2}}\quad \text{and} \quad c_2 = \frac{1}{3\sqrt{2}} \quad & \text{for } \tilde{B}_Q\pi, \\
    c_0 = 0\quad \text{and} \quad c_2 = -\frac{1}{3} \quad & \text{for } \tilde{B}_Q^*\pi,
\end{eqnarray}
while those for $B_Q^*(3/2^-)$ decays to the $\tilde{B}_Q\pi$ channel with $h=1/2$ are given by
\begin{eqnarray}
    c_0 = 0\quad \text{and} \quad c_2 = \frac{1}{3},
\end{eqnarray}
and for the decays to the $\tilde{B}_Q^*\pi$ channel
\begin{eqnarray}
    c_0^{1/2} = -\frac{1}{\sqrt{2}}\quad \text{and} \quad c_2^{1/2} = \frac{\sqrt{2}}{3} & \quad \text{for } h=1/2, \quad \\
    c_0^{3/2} = -\frac{1}{\sqrt{2}}\quad \text{and} \quad c_2^{3/2} = 0 & \quad \text{for } h=3/2.\quad 
\end{eqnarray}

The parameters in the present models include the constituent quark mass and the confinement strength $k$. 
They are given by $m_{u(d)} = 0.35$ GeV, $m_s = 0.55$ GeV, $m_c = 1.5$ GeV, $m_b = 5.0$ GeV, and $k = 0.029$ GeV$^3$. 
These parameters are fixed to produce the orbital excitation of $\omega_\lambda$ around 300 MeV. 
With these parameters, we can compute the wave function parameters $a_\rho$ and $a_\lambda$ that appear in the decay amplitudes. 
For the $\Xi_Q$, we take the average mass of the light up/down and strange quark in the calculation, which is a sufficiently good assumption, especially in the presence of the heavy quark. 
Despite its simplicity, the model can produce a reasonable description of the experimental observations~\cite{Nagahiro:2016nsx}.

Using the helicity amplitudes, we can compute the two-body decay width
\begin{eqnarray}
  \Gamma = \frac{1}{4\pi} \frac{{q}}{2M_i^2} \frac{1}{(2J+1)} \sum_h |A_h|^2
\end{eqnarray}
where $M_i$ and $J$ are the mass and the spin of the parent particle, and $h$ is the helicity. It is worth noting that in this model, the decay of $\tilde{B}^{(*)}_c \to B_c \pi$ overpredicts the experimental data. Empirically, one needs to reduce the coupling $g_q^A \approx 0.65$ for this decay to reproduce the experimental data~\cite{Nagahiro:2016nsx,Arifi:2017sac}. 
We will adopt this for the present calculation, although it has been shown that the origin of such a difference may be from relativistic corrections~\cite{Arifi:2021orx}. 
Although we present the three-body decay width for $B_Q^*(J^-)$, the $\Xi_c(2815) \to \Xi_c^\prime \pi$ decay happen via a two-body process by emitting one pion since $\Xi_c^\prime$ does not subsequently decay via strong interaction.

%--------------------------------------------------------
\begin{figure*}[t]
\centering
\includegraphics[scale=0.55]{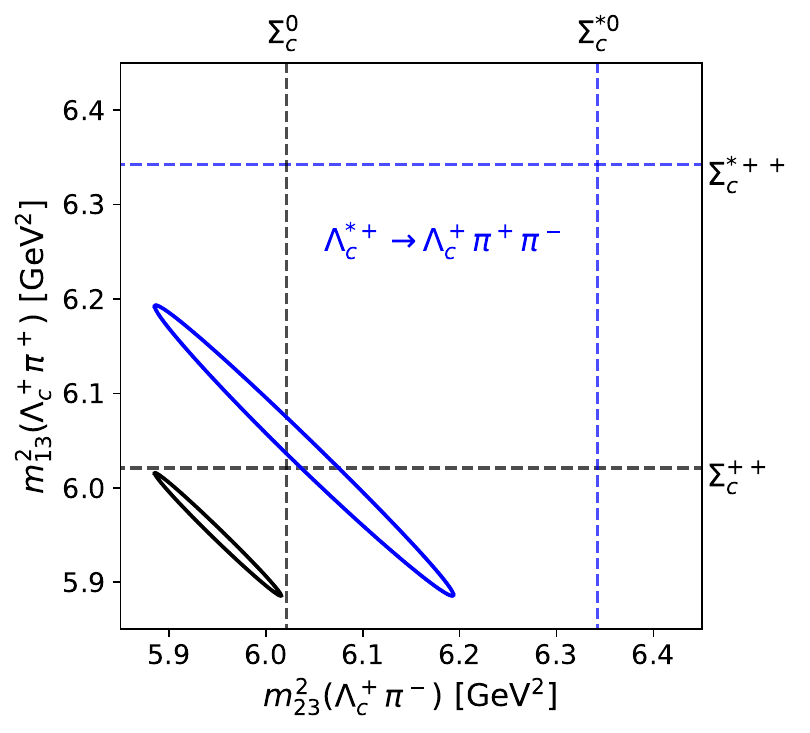}
\includegraphics[scale=0.55]{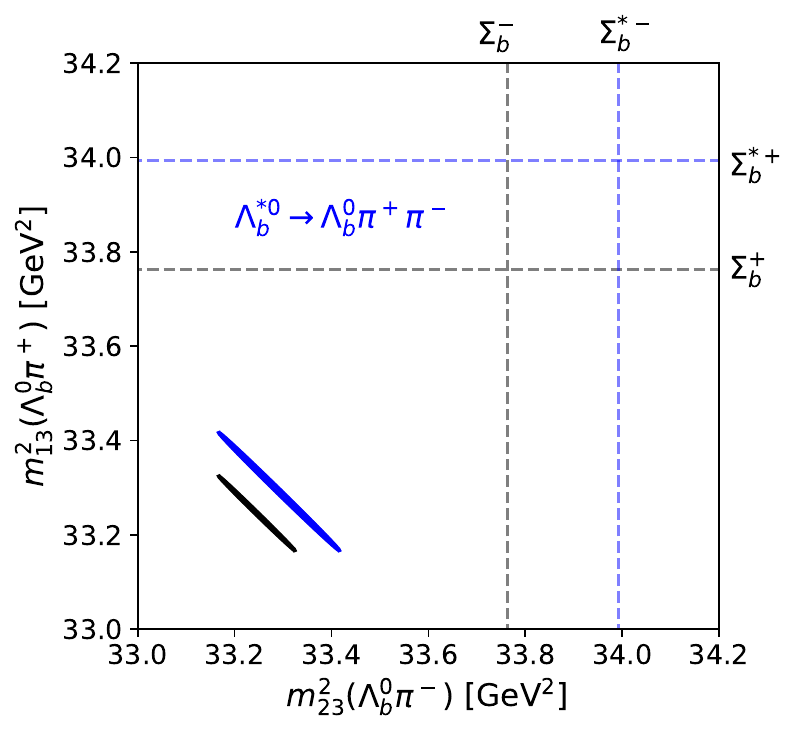}
\includegraphics[scale=0.55]{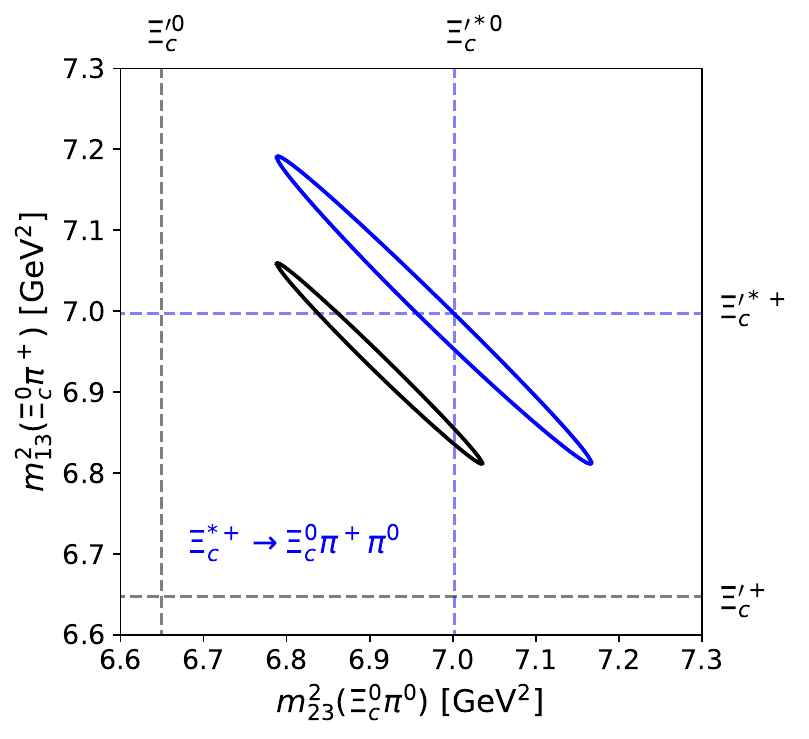}
\includegraphics[scale=0.55]{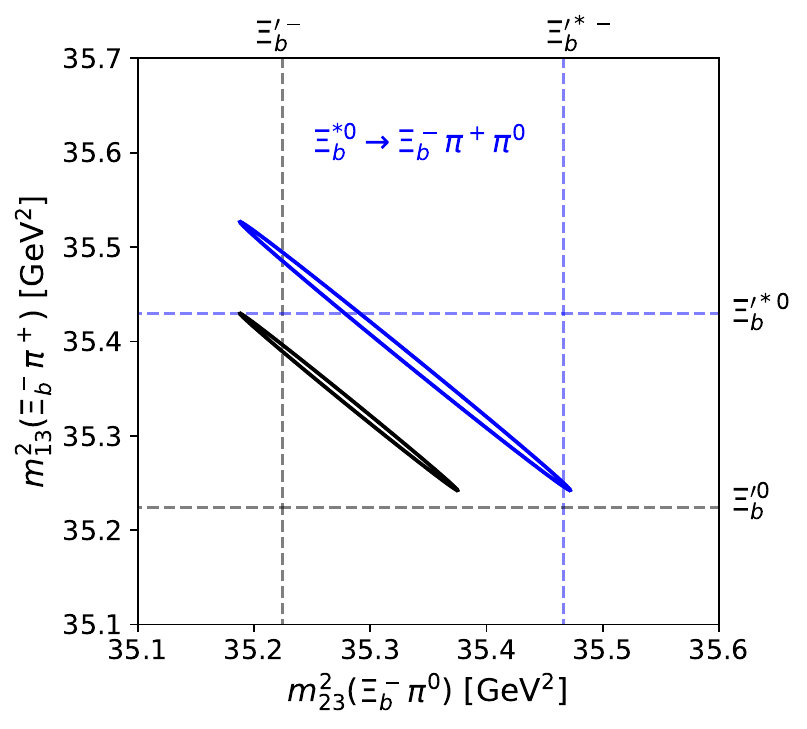}
\caption{\label{fig:boundary} 
Dalitz boundary for two-pion emission decays of $B_Q^*(1/2^-)$ and $B_Q^*(3/2^-)$ with solid lines: (a) $\Lambda_c^*$, (b) $\Xi_c^*$, (c) $\Lambda_b^*$, and (d) $\Xi_b^*$ baryons. The vertical and horizontal resonance bands of $B_Q(1/2^+)$ and $B_Q^*(3/2^+)$ intermediate states are also shown with dashed lines. 
The Dalitz boundary and intermediate state are shown with different color, namely, blue ($J=3/2$) and black $(J=1/2)$.}
\end{figure*}
%--------------------------------------------------------

%--------------------------------------------------------
\subsubsection{Coupling constants}
%--------------------------------------------------------

The coupling constants for each vertex in the sequential decays shown in Fig.~\ref{fig:diagram} are extracted from the chiral quark model~\cite{Arifi:2017sac,Arifi:2018yhr}. Practically, we obtain them by equating the helicity amplitudes in the effective Lagrangian and the quark model as~\cite{Arifi:2017sac} 
\begin{eqnarray}
    A_h^{EL} = A_h^{QM}.
\end{eqnarray}
Following this procedure, the extracted couplings for the second vertex of the sequential decays are expressed as
\begin{eqnarray}
    g_{11}^b &=& \mathcal{O}_1 c_1 \tau,\\
    g_{31}^b &=& - \sqrt{\frac{3}{2}} \mathcal{O}_1 c_1 \tau,
\end{eqnarray}
and those for the first vertex with $B_Q^*(1/2^-)$ are given by
\begin{eqnarray}
    g_{11}^a &=& (\mathcal{O}_0^\lambda c_0 + q^2 \mathcal{O}_2^\lambda c_2) \tau, \\
    g_{13}^a &=& -\sqrt{\frac{3}{2}} \mathcal{O}_2^\lambda c_2 \tau, 
\end{eqnarray}
while those for $B_Q^*(3/2^-)$ are
\begin{eqnarray}
    g_{31}^a &=& -\sqrt{\frac{3}{2}} \mathcal{O}_2^\lambda c_2 \tau,\\
    g_{33}^a &=& \frac{\tau}{2}(\mathcal{O}_0^\lambda (c_0^{1/2}+c_0^{3/2}) + q^2 \mathcal{O}_2^\lambda (c_2^{1/2}+c_2^{3/2})).\quad 
\end{eqnarray}

In two-body decays, the momentum and energy of out-going pion $(\omega, \bm{q})$ are constant. 
However, in three-body decays, we need to replace them with variable energy and momentum of out-going first and second pions $(E_{1}, \bm{p}_{1})$ and $(E_{2}, \bm{p}_{2})$ as functions of invariant masses. 
Consequently, the extracted couplings depend on the variable momentum of the pion for each vertex. 
Care must be taken when incorporating these couplings into the three-body decay amplitudes.

The direct two-pion couplings are estimated from the chiral-partner structure, where they are related to the couplings in the sequential process as~\cite{Kawakami:2018olq,Kawakami:2019hpp}
\begin{eqnarray}
    g_{11}^d = g_{11}^b, \qquad g_{31}^d = g_{31}^b,
\end{eqnarray}
which implies that the initial states of $B_Q^*(3/2^-)$ and $B_Q^*(1/2^-)$ are the chiral partners of the intermediate states $\tilde{B}_Q^*(3/2^+)$ and $\tilde{B}_Q^*(1/2^+)$, respectively, in Fig.~\ref{fig:diagram}, sharing the same coupling.
More details can be found in Ref.~\cite{Arifi:2018yhr}.
Before closing this section, we note again that the coupling constants in the two-pion emission process are fixed from the quark models, not from the fitting to experimental data.

%--------------------------------------------------------
\begin{table*}[t]
\renewcommand{\arraystretch}{1.2}
\caption{Total and partial decay width of $\Lambda_c(2595)^+$ and $\Lambda_c(2625)^+$ decays as well as their bottom counterparts $\Lambda_b(5912)^0$ and $\Lambda_b(5920)^0$. The widths are given in units of MeV, and a comparison with experimental data is also presented. }
\label{tab:Lambda_Q}
\centering
\begin{ruledtabular}
\begin{tabular}{lcccclcccc}
\multirow{2}{*}{Decay mode}	   & \multicolumn{2}{c}{$\Lambda_c(2595)^+$}	 & \multicolumn{2}{c}{$\Lambda_c(2625)^+$} &\multirow{2}{*}{Decay mode}   & \multicolumn{2}{c}{$\Lambda_b(5912)^0$}	 & \multicolumn{2}{c}{$\Lambda_b(5920)^0$} \\ 
 & Our & Expt. & Our & Expt. & & Our & Expt. & Our & Expt.: \\ \hline
(1) $\Lambda_c^+\pi^+\pi^-$& 0.399 &  & 0.325 &  $50.7\%$ & (1) $\Lambda_b^0\pi^+\pi^-$& 0.0031 &  & 0.009 &  \\
$\to \Sigma_c^{0}\pi^+$  &0.182 & 24\% & 0.029 & $5.19\%$ & $\to \Sigma_b^{-}\pi^+$  &0.0005&  & \dots\footref{foot:small} & \\
$\to\Sigma_c^{++}\pi^-$  & 0.162 & 24\% & 0.028 & $5.13\%$ & $\to \Sigma_b^{+}\pi^-$  & 0.0006 &  & \dots\footref{foot:small} & \\
$\to\Sigma_c^{*0}\pi^+$  & \dots\footnote{\label{foot:small}The width is negligibly small due to the kinematically forbidden $D$-wave decay~\cite{Arifi:2017sac}.} &  & 0.043 &  & $\to \Sigma_b^{*-}\pi^+$  & \dots\footref{foot:small} &  & 0.001 & \\
$\to\Sigma_c^{*++}\pi^-$ & \dots\footref{foot:small} &  & 0.044 &  & $\to \Sigma_b^{*+}\pi^-$ & \dots\footref{foot:small} &  & 0.001 & \\
$\to$ Direct (3-body)    & 0.004 & 18\% & 0.053 & Large  & $\to$ Direct (3-body)    & 0.0005 & & 0.002 & \\ 
$\to$ Interference       & 0.054 &  & 0.128 &  & $\to$ Interference       & 0.0015 &  & 0.005 & \\  \hline
(2) $\Lambda_c^+\pi^0\pi^0$& 1.537 &  & 0.245 &  & (2) $\Lambda_b^0\pi^0\pi^0$& 0.0077 &  & 0.013 & \\
$\to\Sigma_c^{+}\pi^0$   & 1.494 &  & 0.037 &  & $\to \Sigma_b^{0}\pi^0$  & 0.0028 &  & \dots\footref{foot:small} & \\
$\to\Sigma_c^{*+}\pi^0$  & \dots\footref{foot:small} &  & 0.071 &  & $\to \Sigma_b^{*0}\pi^0$ & \dots\footref{foot:small} &  & 0.003 & \\
$\to$ Direct (3-body)    & 0.004 &  & 0.039 &  & $\to$ Direct (3-body)    & 0.0012 &  & 0.003 & \\
$\to$ Interference       & 0.038 &  & 0.098 &  & $\to$ Interference       & 0.0036 &  & 0.007 & \\ \hline 
Total          & 1.936 & $2.6\pm0.6$ & 0.570 & $<0.52$ & Total    & 0.0108 & $<0.25$ & 0.022 & $<0.19$ \\
\end{tabular}
\end{ruledtabular}
\renewcommand{\arraystretch}{1}
\end{table*}  
%--------------------------------------------------------

%--------------------------------------------------------
\section{Numerical results and discussion}
%--------------------------------------------------------
\label{sec:result}

In this section, we present numerical findings for the two-pion emission decays of the $\Lambda_Q^*$ and $\Xi_Q^*$ states, with the quark flavor $Q=c$ (charm) or $b$ (bottom), and $J^P=(1/2^-,3/2^-)$. It is notable that we have extensively investigated the $\Lambda_c^*(1/2^-,3/2^-)\to \Lambda_c\pi\pi$ decays in previous studies~\cite{Arifi:2017sac,Arifi:2018yhr}, and now we extend this analysis to encompass $\Lambda_b^*$ and $\Xi_Q$ decays. 
Moreover, our present study involves convoluted Dalitz plots to enable a more realistic comparison with experimental data. 
Additionally, we conduct a comparison with the recent Belle data~\cite{Belle:2022voy} on $\Lambda_c(2625)^+$ to obtain insights into the structure of heavy baryons within the quark model.

%--------------------------------------------------------
\subsection{Dalitz boundaries and resonance bands}
%--------------------------------------------------------

To provide an overview of the decays of $\Lambda_Q^*$ and $\Xi_Q^*$, we present the Dalitz boundaries with possible intermediate states for each decay, as depicted in Fig.~\ref{fig:boundary}. While several decay channels with different charges exist, we illustrate only one representative channel, which is particularly useful before discussing the detailed structure within the Dalitz plot.

In general, we can summarize the decay channels and their partial waves as follows:
\begin{eqnarray}
    B_Q^*(1/2^-) &\to& \tilde{B}_Q(1/2^+) \pi \qquad \text{($S$ wave)},\\
                 &\to& \tilde{B}_Q^*(3/2^+) \pi \qquad \text{($D$ wave)},
\end{eqnarray}
and 
\begin{eqnarray}
    B_Q^*(3/2^-) &\to& \tilde{B}_Q(1/2^+) \pi \qquad \text{($D$ wave)},\\
             &\to& \tilde{B}_Q^*(3/2^+) \pi \qquad \text{($S$ wave)},
\end{eqnarray}
which provide a quick overview of the structure within the plots, as the $S$-wave decay tends to dominate in low-energy reactions.

In Fig.~\ref{fig:boundary}, it is evident that the $\Sigma_c(1/2^+)$ with the $S$ wave resides near the boundary for $\Lambda_c^*(1/2^-)$, while it lies well inside the plot but with the $D$ wave for $\Lambda_c^*(3/2^-)$. 
The vicinity of $\Sigma_c(1/2^+)$ resonance near boundary leads to a well-known isospin breaking effect~\cite{Arifi:2017sac}.
The $\Sigma_c^*(3/2)^+$ lies outside the plots and thus its contribution is negligible for $\Lambda_c^*(1/2^-)$ due to its $D$ wave, although its contribution is suppressed despite the $S$ wave coupling to $\Lambda_c^*(3/2^-)$. 
For $\Lambda_b^*(1/2^-)$ and $\Lambda_b^*(3/2^-)$, the $\Sigma_b^{(*)}$ resonance bands are completely outside the Dalitz plot, resulting in their contributions being much suppressed and appearing as nonresonant structures within the Dalitz plot.

In $\Xi_c^*(J^-)$ decays, the $\Xi^\prime_c(1/2^+)$ resonance is positioned below the Dalitz plot because it lacks the necessary phase space to decay through one-pion emission. 
However, the $\Xi_c^\prime(3/2^+)$ band will be greatly enhanced in $\Xi_c^*(3/2^-)$ decays due to its $S$-wave nature, whereas it decays via $D$ wave in $\Xi_c^*(1/2^-)$, leading to a small branching fraction for the three-body decays. 
For $\Xi_b^*(1/2^-)$, it is shown that the $\Xi_b^{\prime 0}$ is slightly outside the plot of $\Xi_b^-\pi^+\pi^-$ channel as shown in Fig.~\ref{fig:boundary}. 
However, the resonance is well inside the plot for the neutral $\Xi_b^0\pi^0\pi^0$ channel. 
This situation is similar to the $\Lambda_c^*(1/2^-)$ case, which leads to an isospin breaking effect.
A similar situation also applies to the $\Xi_b^*(3/2^-)$ case.
Furthermore, for $\Xi_b^*(3/2^-)$ decays, it is intriguing that $\Xi^\prime_b(1/2^+)$ and $\Xi^\prime_b(3/2^+)$ lie well within the plots, but they have $D$- and $S$-wave couplings, respectively. Meanwhile, only the $\Xi^\prime_b(1/2^+)$ band will be visible in the decay of $\Xi_b^*(1/2^-)$. 
Although we can discuss the contribution from the resonance in Fig.~\ref{fig:boundary}, the role of the direct process cannot be easily shown, and further analysis of the structure of the Dalitz plots will provide insight into that aspect.

%--------------------------------------------------------
\begin{figure*}[t]
	\centering
	\includegraphics[scale=0.5]{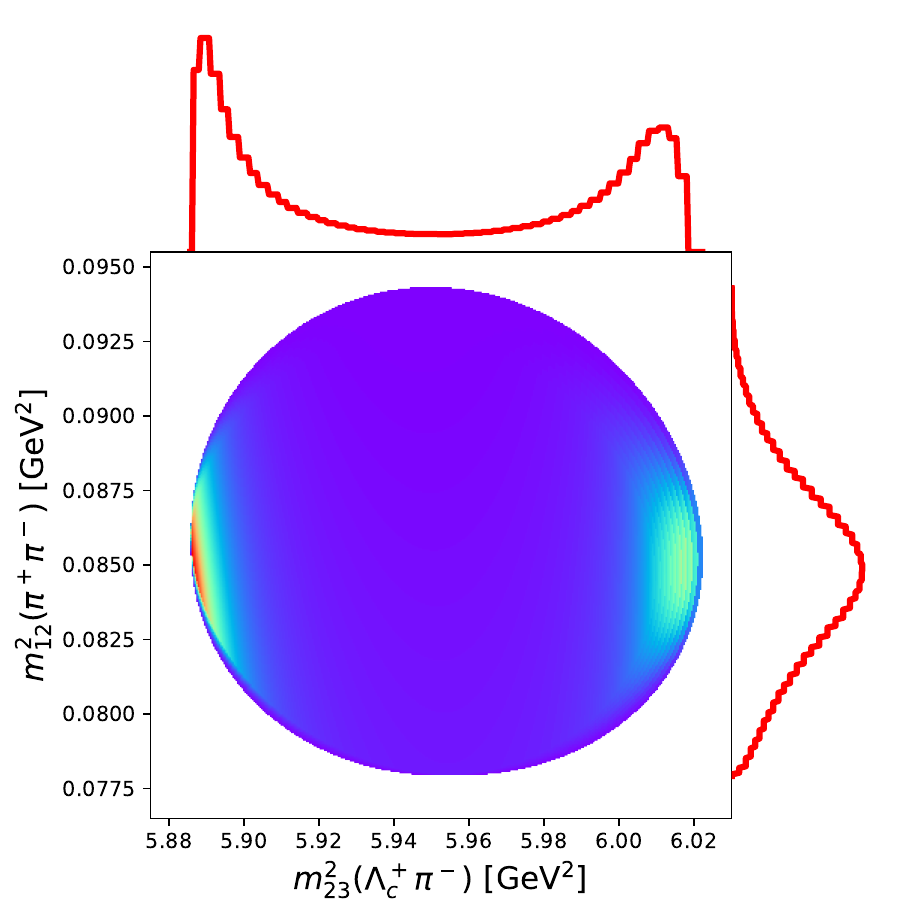}
        \includegraphics[scale=0.615]{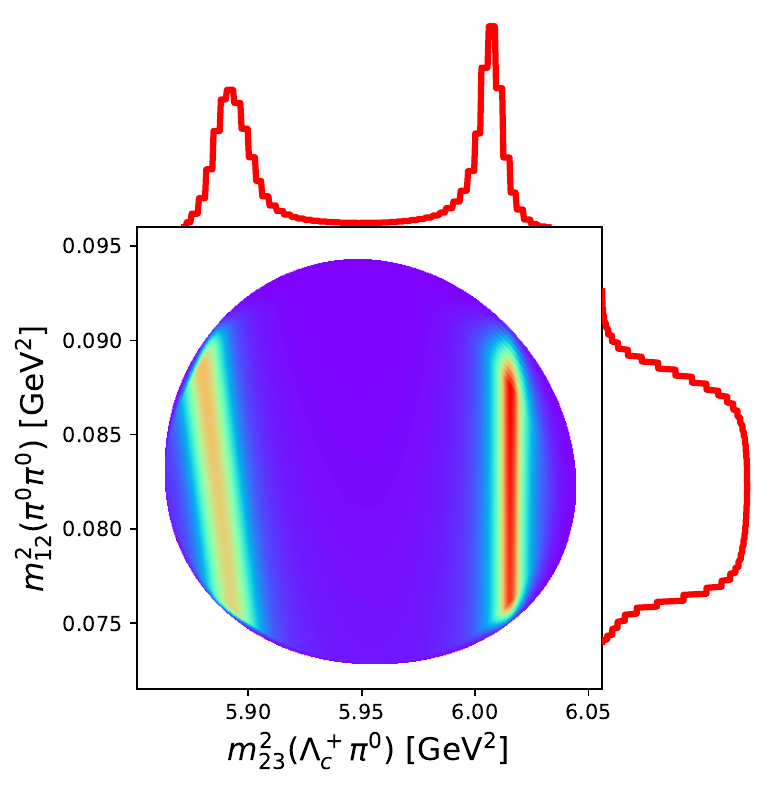}
	\caption{\label{fig:Lc_dalitz12} 
	Dalitz and invariant mass plots of $\Lambda_c(2595)^+$ as a $1/2^-$ state with the $\lambda$ mode in the quark model to (left panel) $\Lambda_c^+\pi^+\pi^-$ and (right panel) $\Lambda_c^+\pi^0\pi^0$. Although the width of  $\Lambda_c(2595)^+$ is around 2 MeV, the smearing effects are clearly visible in the Dalitz plot, making the kinematical reflection and the resonance band near the boundary smeared significantly. }
\end{figure*}
%--------------------------------------------------------

%--------------------------------------------------------
\subsection{ $\Lambda_c^*(1/2^-,3/2^-)$ decays } 
%--------------------------------------------------------

In previous studies~\cite{Arifi:2017sac,Arifi:2018yhr}, we extensively discussed the decays of $\Lambda_c(2595)^+$ and $\Lambda_c(2625)^+$ as $1/2^-$ and $3/2^-$ states, respectively, by considering both $\lambda$ and $\rho$ mode excitations. 
From the mass spectroscopy analysis~\cite{Yoshida:2015tia}, it was known that the $\lambda$ mode is preferable assignment for these states. 
In this discussion, we will provide an analysis of their two-pion emission decay, particularly in light of the new Belle data on $\Lambda_c(2625)^+$~\cite{Belle:2022voy}. 
Additionally, we will include the finite width effect, which smears the Dalitz plot.

%--------------------------------------------------------
\subsubsection{$\Lambda_c(2595)^+$}
%--------------------------------------------------------

Firstly, let us briefly discuss $\Lambda_c(2595)^+$, which is the first excited state of $\Lambda_c^+$. 
In PDG~\cite{ParticleDataGroup:2022pth}, the quantum number $I(J^P)=0(1/2^-)$ is known and it fits with the quark model expectation with the $1/2^-$ state with the $\lambda$ mode from the study of its mass and decay~\cite{Nagahiro:2016nsx,Yoshida:2015tia}.
It is known that the $\rho$-mode assignment for $1/2^-$ state will produce much higher mass and broader decay width, that can be ruled out from the experimental observation.
Due to the vicinity to the $\Sigma_c\pi$ threshold, some non-standard behaviors beyond the three-quark picture are discussed in model calculations~\cite{Nieves:2024dcz,Guo:2016wpy,Lu:2016gev}.
Furthermore, a mixture between the standard state and exotics might also happen in such a case~\cite{Lu:2022puv}.

The decay of $\Lambda_c(2595)^+$ is also known to exhibit a large isospin symmetry breaking~\cite{Arifi:2017sac}. 
As mentioned in PDG~\cite{ParticleDataGroup:2022pth}, the ratio of neutral and charged channels is estimated as
\begin{eqnarray}
    R_{+-}^{00} = \frac{\mathcal{B}(\Lambda_c^+\pi^0\pi^0)}{\mathcal{B}(\Lambda_c^+\pi^+\pi^-)} \approx 4.
\end{eqnarray}
This ratio is inferred from CDF analysis~\cite{CDF:2011zbc} with the measured mass gap of $\Lambda_c(2595)^+$ and $\Lambda_c^+$ in PDG~\cite{ParticleDataGroup:2022pth}.
In the Table.~\ref{tab:Lambda_Q}, we show that a simple quark model gives $R_{+-}^{00}=1.537/0.399=3.85$, which is consistent with the PDG estimate.
Recently, BESIII measured the absolute branching fraction of $\Lambda_c^+\pi^+\pi^-$ channel, but only the upper limit $\mathcal{B}<80.8\%$ is obtained~\cite{BESIII:2024udz} due to the limited statistics. 
In PDG, its branching fraction is $66\%$ [$\Sigma_c^0\pi^+$(24\%),$\Sigma_c^{++}\pi^-$(24\%),3-body(18\%)], still assuming the isospin symmetry and show different result from our model expectation in Table~\ref{tab:Lambda_Q}.
However, because of the dominant neutral decay channel, the branching fraction of charged channel is much suppressed and it can be as small as $\mathcal{B}\approx 20\%$ in our estimate.
Further experimental analysis is necessary to improve the current situation.

%--------------------------------------------------------
\begin{figure}[t]
	\centering
    \includegraphics[scale=0.55]{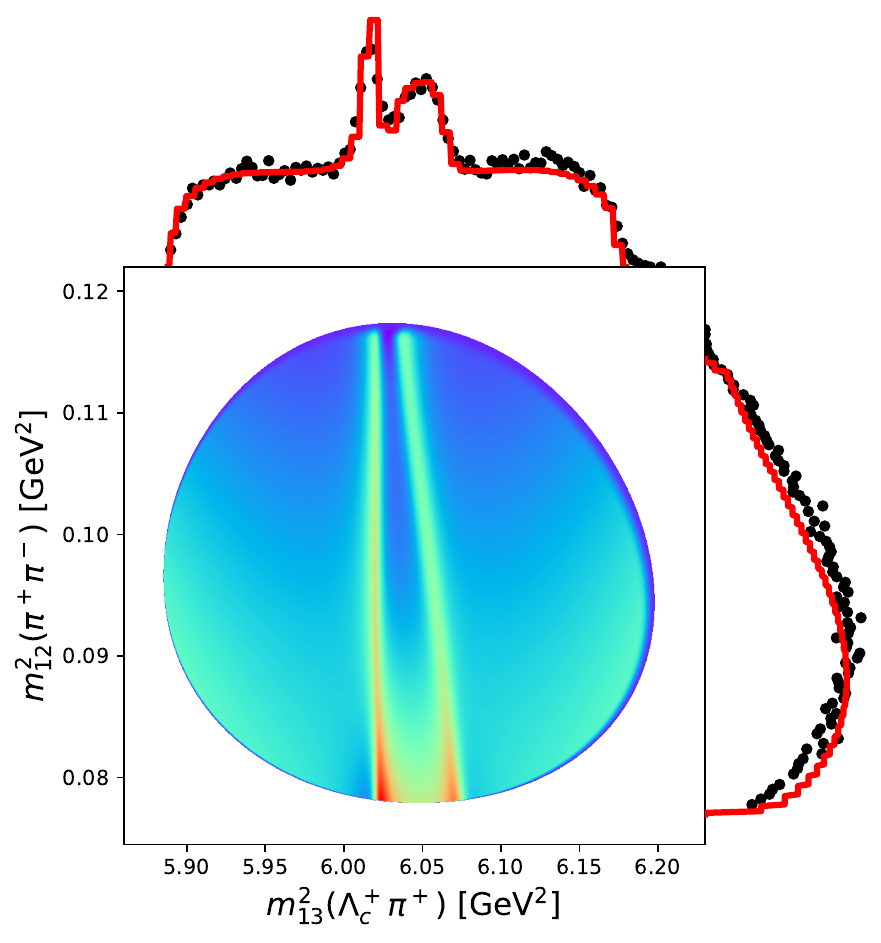}
	\caption{\label{fig:Lc_dalitz32} 
	Dalitz and invariant mass plots of $\Lambda_c(2625)^+$ decaying into $\Lambda_c^+\pi^+\pi^-$, assigned as a $3/2^-$ state with the $\lambda$ mode. Comparison with recent Belle data~\cite{Belle:2022voy} for the invariant mass distribution is shown.}
\end{figure}
%--------------------------------------------------------
%--------------------------------------------------------
\begin{figure}[t]
	\centering
        \includegraphics[scale=0.5]{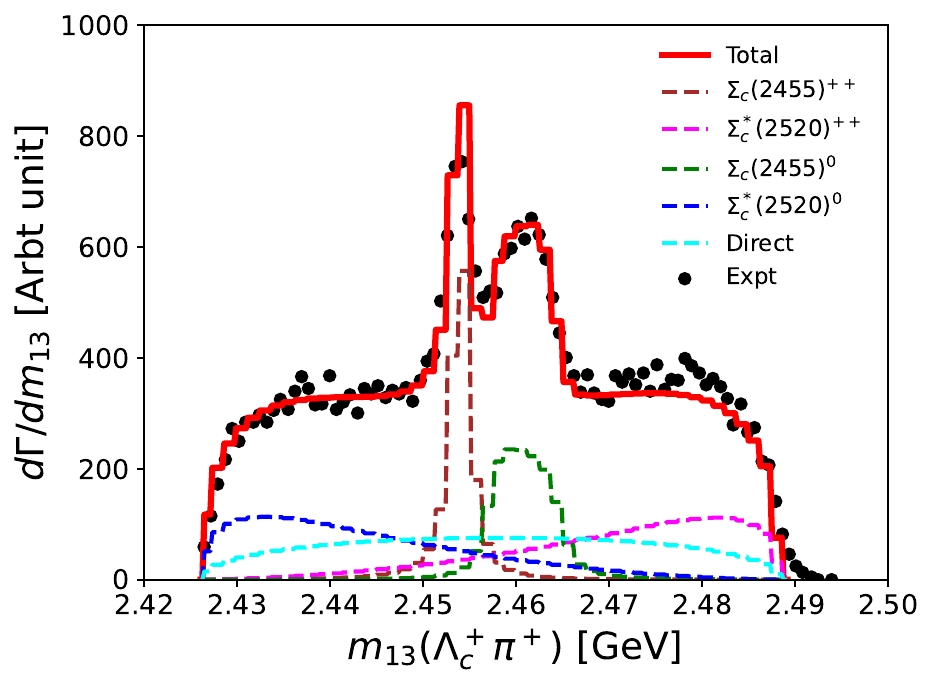}
        \includegraphics[scale=0.5]{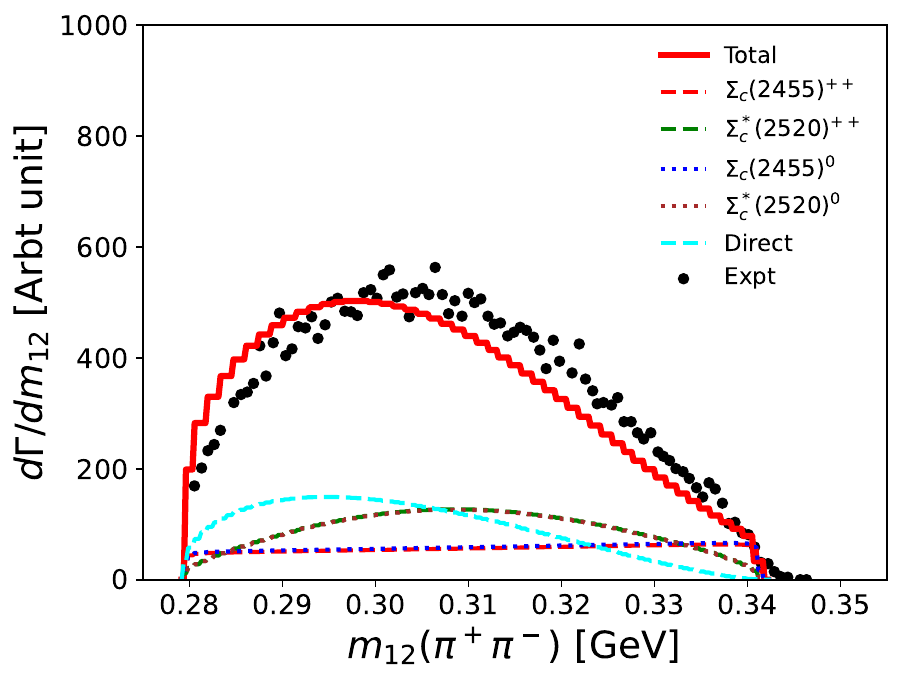}
	\caption{\label{fig:Lc_inv} 
    The ${\Lambda_c^+\pi^+}$ and ${\pi^+\pi^-}$ invariant mass plots of $\Lambda_c(2625)^+ \to\Lambda_c^+\pi^+\pi^-$, along with their components, compared with recent Belle data~\cite{Belle:2022voy}. The components for the ${\pi^+\pi^-}$ invariant mass plot are doubled to better illustrate their shape. }
\end{figure}
%--------------------------------------------------------

%--------------------------------------------------------
\begin{figure*}[t]
	\centering
    \includegraphics[scale=0.55]{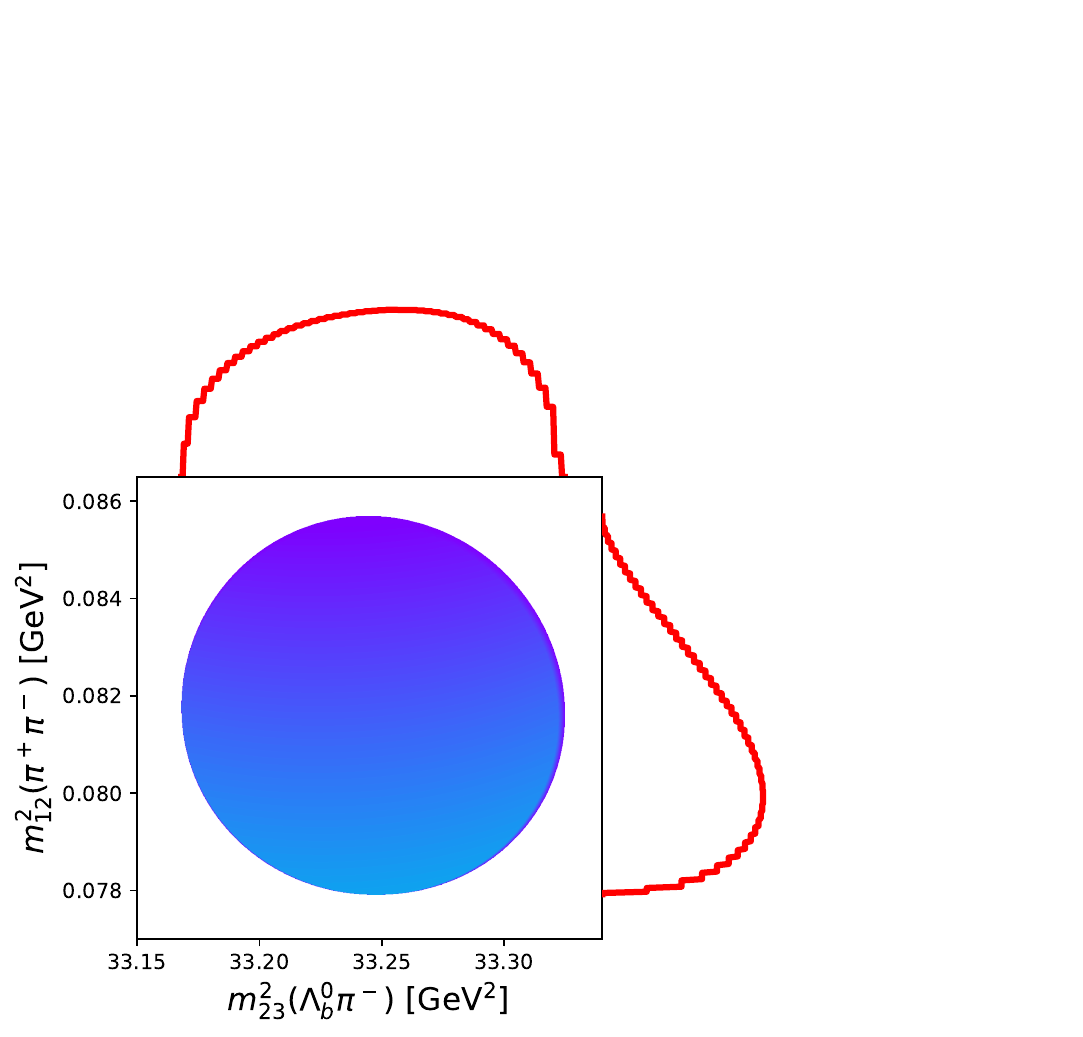}
    \includegraphics[scale=0.55]{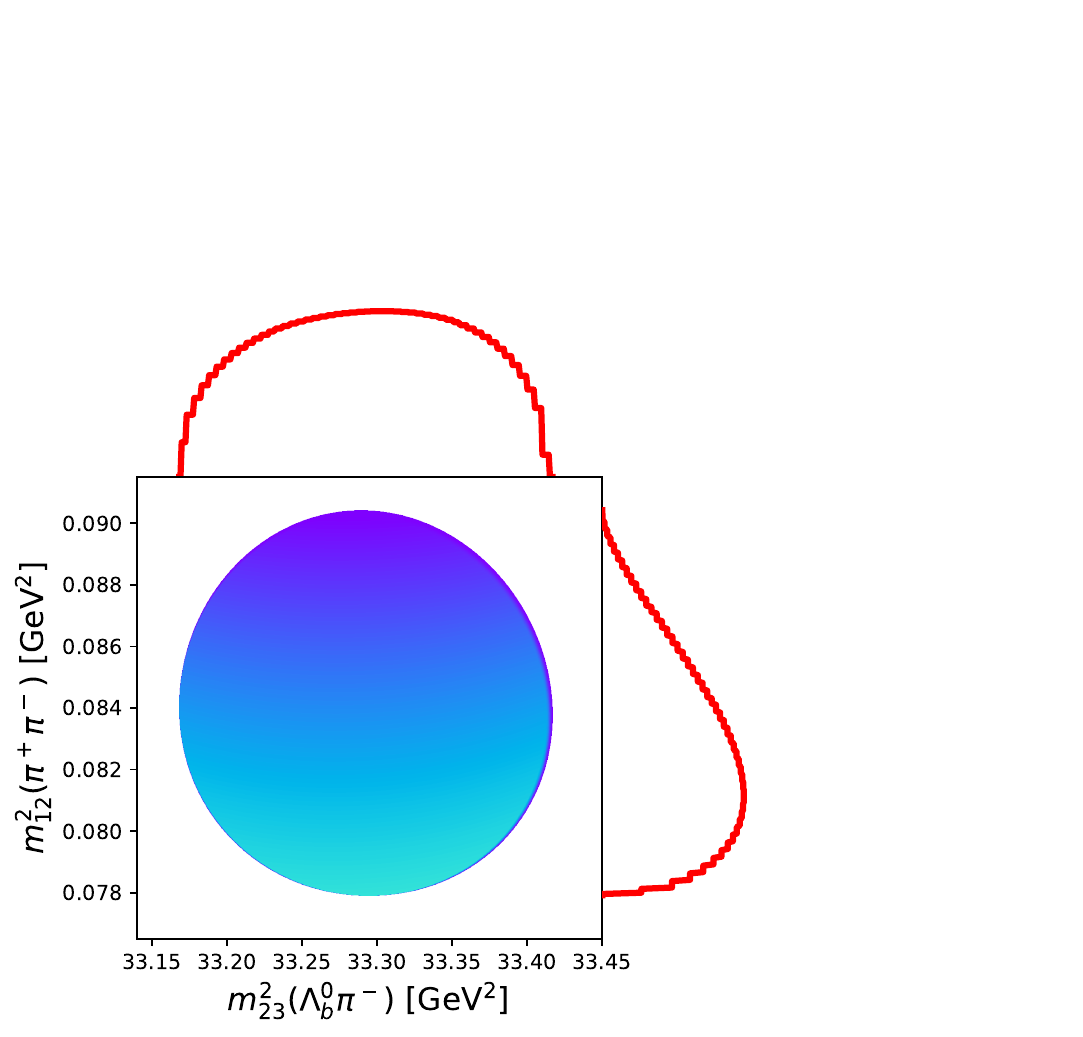}
	\caption{\label{fig:Lb_dalitz} 
	Dalitz and invariant mass plots of (a) $\Lambda_b(5912)^0$ and (b) $\Lambda_b(5920)^0$ to $\Lambda_b^0\pi^+\pi^-$ assigned as $1/2^-$ and $3/2^-$ states with the $\lambda$ mode, respectively. The decay is dominated by the non-resonant process and no explicit $\Sigma_b^{(*)}$ resonance bands are observed in the plot. }
\end{figure*}
%--------------------------------------------------------

Fig.~\ref{fig:Lc_dalitz12} shows some results of the Dalitz plot for $\Lambda_c(2595)^+\to\Lambda_c^+\pi^+\pi^-$ and $\Lambda_c^+\pi^0\pi^0$, which different positions of $\Sigma_c(1/2^+)$ resonance bands.
It is seen that the direct process is small and hindered by the dominant $S$-wave resonance.
Although the width is around 2 MeV, the finite width effect is visible where the resonance bands near the Dalitz boundary and the kinematical reflection (diagonal bands) are smeared.
Such an effect, which give differences with those shown in previous works~\cite{Arifi:2017sac,Arifi:2018yhr}, provides a more realistic Dalitz plot.
At the moment, there is no available data for the Dalitz plot of $\Lambda_c(2595)^+$, but we expect that the comparison of those observables can constrain its quark structure and clarify the threshold effect.

%--------------------------------------------------------
\subsubsection{$\Lambda_c(2625)^+$}
%--------------------------------------------------------

The $\Lambda_c(2625)^+$ is the second excited state of $\Lambda_c^+$ and its quantum number $I(J^P)=0(3/2^-)$ is displayed in PDG~\cite{ParticleDataGroup:2022pth}.
The spin and parity are yet to be measured, but in the simple quark model~\cite{Yoshida:2015tia,Nagahiro:2016nsx} it is naturally assigned as a $3/2^-$ state with the $\lambda$ mode and as a HQS partner of $\Lambda_c(2595)^+$.
However, there are several recent development in experimental side~\cite{Belle:2022voy,BESIII:2024udz} which further constrain the internal structure of $\Lambda_c(2595)^+$.

The absolute branching ratio $\mathcal{B}(\Lambda_c^+\pi^+\pi^-)$ is recently measured to be 50.7\% by BESIII~\cite{BESIII:2024udz}, slightly smaller than the isospin symmetric expectation $\mathcal{B}=67\%$.
Our model also predict the value $\mathcal{B}=(0.325/0.570)100\%=57\%$, which does not deviate from the data and isospin symmetry very much.
The upper limit of the decay width $\Gamma_\text{exp} < 0.52$ MeV is improved by Belle~\cite{Belle:2022voy} and our calculation $\Gamma_\text{our}= 0.57$ MeV show good agreement with the data as shown in Table~\ref{tab:Lambda_Q}.
The branching fractions are also measured~\cite{Belle:2022voy} 
\begin{eqnarray}
   R_1^0 &=& \frac{\mathcal{B}(\Sigma_c^0\pi^+)}{\mathcal{B}( \Lambda_c^+\pi^+\pi^-)} = (5.19 \pm 0.23)\%, \\
   R_2^{++} &=&  \frac{\mathcal{B}(\Sigma_c^{++}\pi^-)}{\mathcal{B}(\Lambda_c^+\pi^+\pi^-)} = (5.13 \pm 0.26)\%.\quad\quad 
\end{eqnarray}
Our results are a slightly larger than the data of around $R_1^0=8.92\%$ and $R_2^{++}=8.61\%$.\footnote{We find that there was an inaccuracy in the previous works~\cite{Arifi:2017sac,Arifi:2018yhr} where the ratio for the cross diagram in Fig.~\ref{fig:baryon} is much larger due to the incorrect momentum assignment in the coupling constants.} 
In Belle analysis~\cite{Belle:2022voy}, there is a phasespace (PHSP) background that is not taken into account in our calculation, which may result in the discrepancy.
Also, one can improve the prediction by adjusting the model parameters.
However, here we do not fit the data as our purpose is to provide a comparison with the simple quark model.

In Fig.~\ref{fig:Lc_dalitz32}, we present our calculated Dalitz and invariant mass plots, which agree with those reported by Belle~\cite{Belle:2022voy}. 
Since the width is small $\Gamma_\mathrm{exp}<0.52$ MeV, the smearing effect is rather small, namely, only give a small enlargement of the plot as seen near the boundary.
In the upper and right sides of the Dalitz plot, we compare the structure of the ${\Lambda_c^+\pi^+}$ and ${\pi^+\pi^-}$ invariant mass distributions with recent Belle data~\cite{Belle:2022voy}. 
Moreover, we show the components of each amplitude to the invariant mass plot in Fig.~\ref{fig:Lc_inv}.
Note that this is a simple comparison with the data, not obtained by the fit.
The coupling constants are computed from the quark model~\cite{Nagahiro:2016nsx} and their ratio of the couplings is important in determining the structure in invariant mass plot.

As shown in Fig.~\ref{fig:Lc_inv}, the invariant mass distributions are reasonably reproduced by assigning $\Lambda_c(2625)^+$ as a $3/2^-$ state with the $\lambda$ mode, not only the shape of the two peaks and shoulders in the ${\Lambda_c^+\pi^+}$ invariant mass, but also the asymmetric shape of the ${\pi^+\pi^-}$ invariant mass distribution. 
Note that the shoulder will be much larger than expected for the $\rho$-mode excitation as shown in previous study~\cite{Arifi:2017sac,Arifi:2018yhr} that can be ruled out from the Belle data~\cite{Belle:2022voy}.
The results of components seem to be consistent, but the $\Sigma_c(1/2^+)$ contribution is larger than the Belle data, with in accordance with the branching fraction $R_1^0$ and $R_2^{++}$ discussed previously.
Furthermore, the asymmetry in the ${\pi^+\pi^-}$ invariant mass can be well described where the direct two-pion coupling constants are estimated by assuming the chiral-partner structure~\cite{Kawakami:2018olq}.

%--------------------------------------------------------
\subsection{ $\Lambda_b^*(1/2^-,3/2^-)$ decays } 
%--------------------------------------------------------

Next, we will discuss the bottom counterparts of $\Lambda_c^*(1/2^-,3/2^-)$. The situation is rather different since there is no explicit resonance band in the Dalitz plot, resulting in a small decay width. 
We assign the $\Lambda_b(5912)^0$ and $\Lambda_b(5920)^0$ as $1/2^-$ and $3/2^-$ states, respectively, with the $\lambda$ mode.

%--------------------------------------------------------
\subsubsection{ $\Lambda_b(5912)^0$} 
%--------------------------------------------------------

The $\Lambda_b(5912)^0$ represents the first excited state of bottom baryons observed in LHCb~\cite{LHCb:2012kxf,LHCb:2020lzx} and CMS~\cite{CMS:2020zzv}. 
In PDG~\cite{ParticleDataGroup:2022pth}, it is expected to have the spin-parity $J^P=1/2^-$ with a decay width $\Gamma_{\rm exp} < 0.66$ MeV and a dominant $\Lambda_b\pi\pi$ mode. 
Unlike the charm sector, where the $\Lambda_c(2595)^+$ lies very close to threshold, resulting in a significant isospin symmetry breaking, the $\Lambda_b(5912)^0$ baryon cannot decay into $\Sigma_b(1/2^+)\pi$ or $\Sigma_b^*(3/2^+)\pi$ via one-pion emission as implied in Fig.~\ref{fig:boundary}. 
Consequently, its primary decay mode is the direct three-body decay process into $\Lambda_b\pi\pi$, leading to a very narrow decay width.

In the quark model, the calculated decay width with the $\lambda$ mode is consistent with the upper limit of the experimental data as shown in Table~\ref{tab:Lambda_Q}.
Since the two-body decays of $\Lambda_b(5912)^0$ into $\Sigma_b\pi$ and $\Sigma_b^*\pi$ are kinematically forbidden, we consider a three-body decay through $\Sigma_b^{(*)}$ in the intermediate states which contribute virtually.
The main contribution is mainly coming from the $\Sigma_b\pi$ channel with the $S$ wave.
We observe that the direct process is significant partly because the $\Sigma_b\pi$ decay is not open and thus suppressed. 
The Dalitz and invariant mass plot of $\Lambda_b(5912)^0$ is shown in the left panel of Fig.~\ref{fig:Lb_dalitz}, showing only a non-resonant contribution. 
Similar to the $\Lambda_c(2625)^+$ case, the asymmetry in the $\pi^+\pi^-$ is observed due to the present of the direct process~\cite{Kawakami:2018olq}.

%--------------------------------------------------------
\begin{figure}[b]
	\centering
    \includegraphics[scale=0.5]{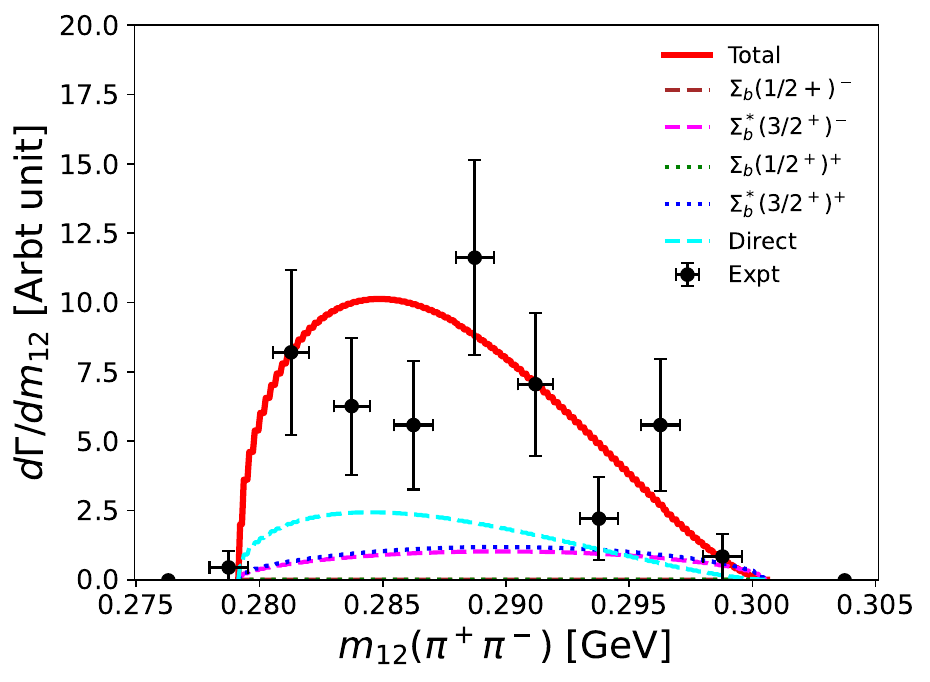}
	\caption{\label{fig:Lb_inv} 
    The ${\pi^+\pi^-}$ invariant mass plot of the $\Lambda_b(5920)^0\to \Lambda_b^0\pi^+\pi^-$ decay, compared with existing LHCb data~\cite{LHCb:2012kxf}.
    Additional statistics are required to further validate the asymmetry predicted by our model.}
\end{figure}
%--------------------------------------------------------

%--------------------------------------------------------
\subsubsection{ $\Lambda_b(5920)^0$}
%--------------------------------------------------------

The $\Lambda_b(5920)^{0}$ is also a very narrow state with $\Gamma_{\rm exp} < 0.63$ MeV and the spin-parity $J^P=3/2^-$ is assigned in PDG~\cite{ParticleDataGroup:2022pth}.
This $\Lambda_b(5920)^{0}$ was observed in LHCb~\cite{LHCb:2012kxf,LHCb:2020lzx}, CDF~\cite{CDF:2013pvu} and CMS~\cite{CMS:2020zzv}. 
Along with the $\Lambda_b(5912)^0$, the $\Lambda_b(5920)^0$ forms a HQS doublet with their mass difference is around 8 MeV.
The $\Lambda_b(5920)^{0}$ mainly decays into $\Lambda_b\pi\pi$, and the non-resonant process contributes significantly to this decay process since the threshold of $\Sigma_b^{(*)}\pi$ decay is not open as shown in Fig.~\ref{fig:boundary}. 
However, the $\Sigma_b^{(*)}$ resonance can contribute virtually in the three-body decay.
Opposite to the $\Lambda_b(5920)^0$, the $\Sigma_b^*$ resonance has a dominant contribution due to the $S$-wave nature, while the $\Sigma_b$ contribution is negligible because of the $D$-wave nature.
In Table~\ref{tab:Lambda_Q}, we present the $\lambda$-mode excitation and the calculated decay width is consistent with the upper limit of the experimental data. 
It is evident that the contributions from the sequential and direct processes are comparable. 
Again, the direct process will be visible when the $S$-wave resonance contribution is suppressed. 

%--------------------------------------------------------
\begin{table*}[t]
\renewcommand{\arraystretch}{1.2}
\caption{Total and partial decay widths of $\Xi_c(2790)^+$ and $\Xi_c(2815)^+$ decays, along with their bottom counterparts $\Xi_b(6087)^0$ and $\Xi_b(6095)^0$. The widths are given in units of MeV, and a comparison with experimental data is also presented.}
\label{tab:Xi_Q}
\centering
\begin{ruledtabular}
\begin{tabular}{lrrrrlrrrr} 
\multirow{2}{*}{Decay mode}	   & \multicolumn{2}{c}{$\Xi_c(2790)^+$}	 & \multicolumn{2}{c}{$\Xi_c(2815)^+$} &\multirow{2}{*}{Decay mode}   & \multicolumn{2}{c}{$\Xi_b(6087)^0$}	 & \multicolumn{2}{c}{$\Xi_b(6095)^0$}  \\ 
 & Our & Expt. & Our & Expt. & & Our & Expt. & Our & Expt. \\ \hline
(1) $\Xi_c^{\prime +}\pi^0$ & 2.3367 & & 0.118 &  & (1) $\Xi_b^0\pi^-\pi^+$  & 0.7440  & & 0.1200  & \\
(2) $\Xi_c^{\prime 0}\pi^+$ & 4.6244 & & 0.215 &  & $\to \Xi_b^{\prime -}\pi^+$ & 0.7467 & &  0.0024 &\\
(3) $\Xi_c^+\pi^-\pi^+$    & 0.0101 &  & 1.460 &  & $\to \Xi_b^{\prime *-}\pi^+$  & \dots\footnote{\label{foot:small}The width is negligibly small due to the kinematically forbidden $D$-wave decay.} & & 0.1056   & \\
$\to \Xi_c^{\prime *0}\pi^+$ & 0.0002 &  & 1.453 &  &  $\to$ Direct (3-body)  & 0.0005 & & 0.0016  &   \\
$\to$ Direct (3-body)  & 0.0098 &  & 0.038 & & $\to$ Interference  & $-0.0032$ & & 0.0103   &    \\
$\to$ Interference & 0.0001 &  & $-0.031$ &  & (2) $\Xi_b^0 \pi^0 \pi^0$ & 0.7070 & &  0.2312  &   \\
(4) $\Xi_c^+\pi^0\pi^0$    & 0.0045 &  & 0.472 &  &$\to \Xi_b^{\prime 0}\pi^0$  & 0.7092  & & 0.0032   &  \\
$\to \Xi_c^{\prime *+}\pi^0$ & 0.0003 &  &0.475 & & $\to \Xi_b^{\prime *0}\pi^0$ & \dots\footref{foot:small} & & 0.2252  &  \\  
$\to$ Direct (3-body)  & 0.0041 &  & 0.014 &  &   $\to$ Direct (3-body) & 0.0004 & &0.0010   & \\
$\to$ Interference & 0.0001&  & $-0.016$ &  & $\to$ Interference & $-0.0027$ & &  0.0017   &    \\
(5) $\Xi_c^0\pi^+\pi^0$    & 0.0114 &  & 1.856 &  &   (3) $\Xi_b^- \pi^+ \pi^0$   & 0.5623 & & 0.2189  &\\
$\to \Xi_c^{\prime *+}\pi^0$ & 0.0002 &  & 0.363 & & $\to \Xi_b^{\prime 0}\pi^0$ & 0.0008 & &  \dots\footref{foot:small}  & \\ 
$\to \Xi_c^{\prime *0}\pi^+$ & 0.0003 &  & 1.507 &  &  $\to \Xi_b^{\prime -}\pi^+$ &  0.5648 & &0.0018     \\
$\to$ Direct (3-body)  & 0.0108 & & 0.041 &  & $\to \Xi_b^{\prime *0}\pi^0$ &  \dots\footref{foot:small} & & 0.1095 &  \\
$\to$ Interference & 0.0002& & $-0.055$ &  & $\to \Xi_b^{\prime *-}\pi^+$ & \dots\footref{foot:small} & &  0.0973  &  \\
&  &  &  &  & $\to$ Direct (3-body) & 0.0004 & &  0.0015   & \\
&  &  &  &  & $\to$ Interference & $-0.0037$ & &0.0088  & \\ \hline 
Total    & 6.9871  & $8.9\pm 1.0$  & 4.121 & $2.43 \pm 0.26$   & Total  & 2.0133 & $2.43\pm 0.51 $  & 0.5701  & $0.50\pm 0.33$ \\
\end{tabular}
\end{ruledtabular}
\renewcommand{\arraystretch}{1}
\end{table*}  
%--------------------------------------------------------

Shown in the right panel of Fig.~\ref{fig:Lb_dalitz} is the Dalitz and invariant mass plots for the $\Lambda_b(5920)^0$ decay, which display similar structure with those in the $\Lambda_b(5912)^0$ decay.
Since the direct process contributes significantly, its $P$-wave nature is observed in the Dalitz plot as an asymmetric pattern. 
Consequently, the $\pi^+\pi^-$ invariant mass distribution is more enhanced in the lower mass region, exhibiting an asymmetric pattern, as seen in Fig.~\ref{fig:Lb_inv}. 
Meanwhile, the $\Lambda_b^0\pi^-$ invariant mass distribution remains symmetric. 
Comparison with LHCb data is made, but more statistics are needed to confirm the asymmetric pattern. 
This phenomenon is similar to the $\Lambda_c(2625)^+$ case and can provide a test of the chiral-partner structure in heavy baryons~\cite{Kawakami:2018olq}.

%--------------------------------------------------------
\subsection{$\Xi_c^*(1/2^-,3/2^-)$ decays}
%--------------------------------------------------------

Although the $\Xi_c^*(1/2^-, 3/2^-)$ states form a doublet with charged and neutral states, we present results only for the charged states here as we expect the structure to be similar. In the quark model, we assign the observed $\Xi_c(2790)^+$ and $\Xi_c(2815)^+$ as a HQS with the $\Xi_c^*(1/2^-)$ and $\Xi_c^*(3/2^-)$ states, respectively.

%--------------------------------------------------------
\subsubsection{$\Xi_c(2790)^+$}
%--------------------------------------------------------

The $\Xi_c(2790)^+$ was observed in $\Xi_c^\prime(1/2^+)\pi$ and $\Xi_c\gamma$ decays, with a total width of $8.9\pm 1.0$ MeV. While the decay $\Xi_c\pi\pi$ has been analyzed by Belle, no significant peak for $\Xi_c(2790)^+$ was found~\cite{Belle:2008yxs,Belle:2016lhy}. In PDG, $\Xi_c(2790)^+$ is assigned as a $1/2^-$ state from the quark model expectation although the spin-parity has not been directly measured. 
The decay of this $\Xi_c(2790)^+$ is consistent with the $1/2^-$ and $\lambda$ mode, as shown in Table~\ref{tab:Xi_Q}. 
Although $\Xi_c(2790)^+ \to \Xi_c\pi\pi$ is also possible with $\Xi_c(2645)$ as an intermediate state, this two-pion emission decay is highly suppressed due to its $D$-wave nature and has not yet been observed experimentally.

%--------------------------------------------------------
\begin{figure*}[t]
\centering
\includegraphics[width=2\columnwidth]{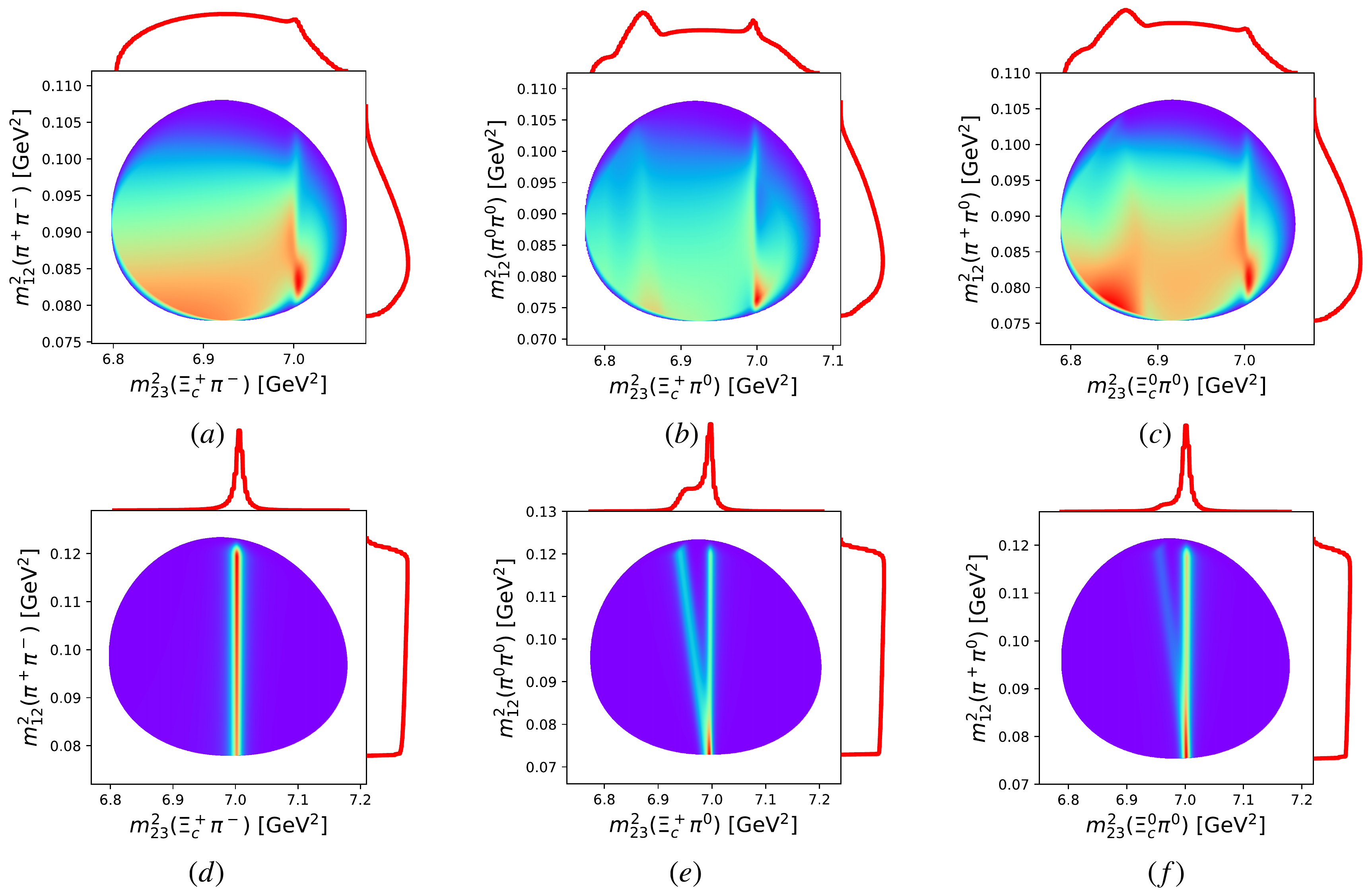}
\caption{\label{fig:Xi_c} 
Dalitz and invariant mass plots of $\Xi_c(2790)^+$ to (a) $\Xi_c^+\pi^+\pi^-$, (b) $\Xi_c^+\pi^0\pi^0$, (c) $\Xi_c^0\pi^+\pi^0$, and $\Xi_c(2815)^+$ to (d) $\Xi_c^+\pi^+\pi^-$, (e) $\Xi_c^+\pi^0\pi^0$, (f) $\Xi_c^0\pi^+\pi^0$. The relative strength of the diagonal resonance bands varies depending on the decay modes.
The direct (nonresonant) process is dominant for the $\Xi_c(2790)^+$ decay whereas it is negligible for the case of $\Xi_c(2815)^+$.}
\end{figure*}
%--------------------------------------------------------

%--------------------------------------------------------
\begin{figure*}[t]
\centering
\includegraphics[width=2\columnwidth]{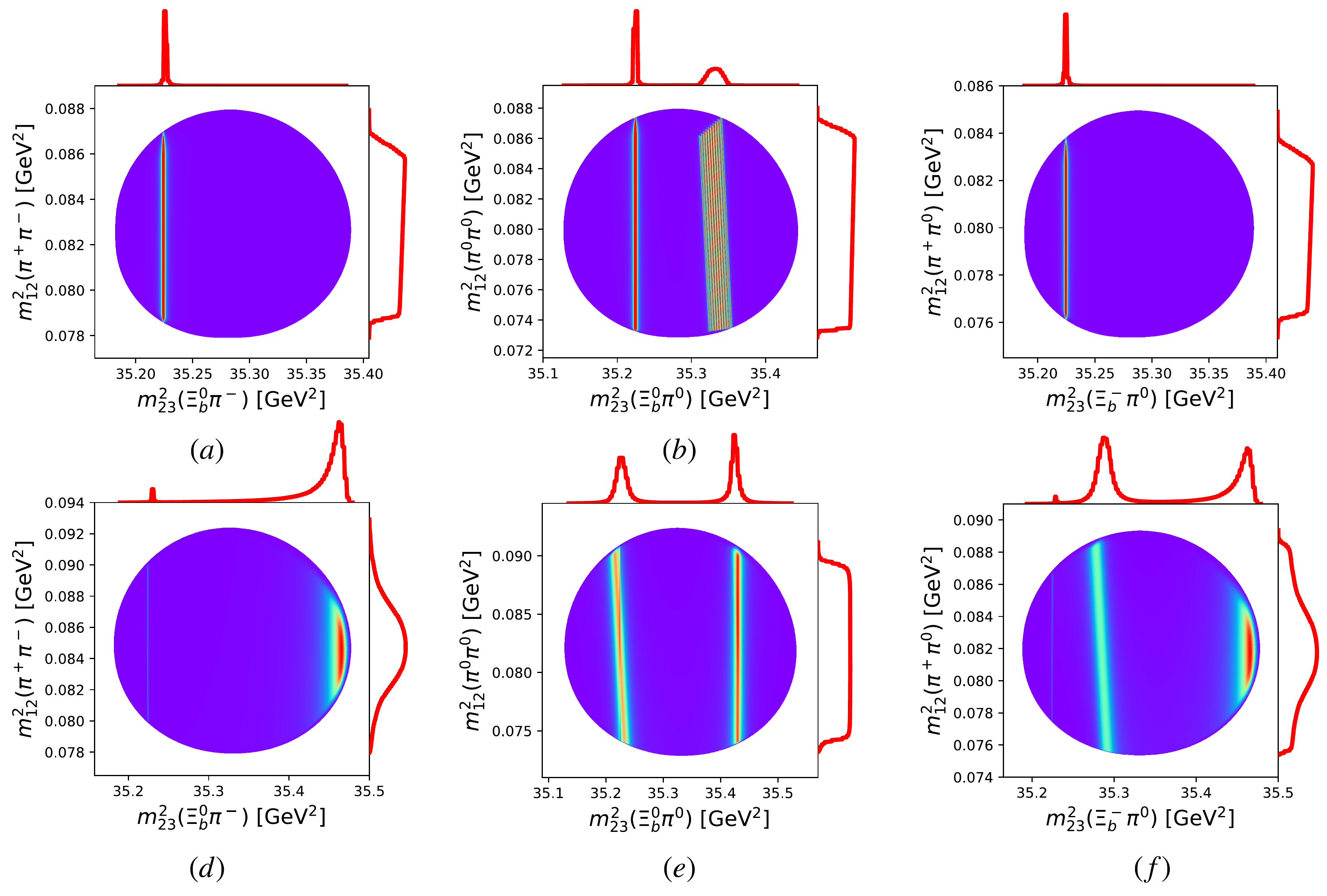}
\caption{\label{fig:Xi_b} 
Dalitz and invariant mass distribution plots of $\Xi_b(6087)^0$ decay to (a) $\Xi_b^0\pi^+\pi^-$, (b) $\Xi_b^0\pi^0\pi^0$, and (c) $\Xi_b^-\pi^+\pi^0$, as well as $\Xi_b(6095)^0$ decay to (d) $\Xi_b^0\pi^+\pi^-$, (e) $\Xi_b^0\pi^0\pi^0$, and (f) $\Xi_b^-\pi^+\pi^0$. The relative strength of the diagonal resonance bands varies depending on the decay modes. }
\end{figure*}
%--------------------------------------------------------

The Dalitz and invariant mass plots of $\Xi_c(2790)^+ \to \Xi_c\pi\pi$ are presented in the upper panel of Fig.~\ref{fig:Xi_c}. 
This plot illustrates various decay modes with different charges: (a) $\Xi_c^+\pi^+\pi^-$, (b) $\Xi_c^+\pi^0\pi^0$, and (c) $\Xi_c^0\pi^+\pi^0$.
Notably, the $\Xi_c'(1/2^+)$ cannot decay through one-pion emission due to limited phase space, so it is absent from the Dalitz plot, and only the $\Xi_c'(2645)$ resonance is visible. Since there is no $S$-wave resonance in the plot, the direct process appears to dominate this decay and the asymmetric shape in the $\pi\pi$ invariant mass is seen, as depicted in Fig. \ref{fig:Xi_c}, which can be tested in experiments. 
The interference also produce the nontrivial structure along the $\Xi_c'(2645)$ resonance band.
However, it may present some challenges due to the small branching fraction to the $\Xi_c \pi \pi$ channel.
We also observe that the smearing effect becomes quite pronounced because the width is approximately 9 MeV. This broadens the Dalitz plot, mainly impacting the upper right region.

%--------------------------------------------------------
\subsubsection{$\Xi_c(2815)^+$}
%--------------------------------------------------------

The $\Xi_c(2815)^+$ was observed in $\Xi_c\pi\pi$, $\Xi_c^\prime\pi$, and $\Xi_c\gamma$ decays, with a measured total width of $2.43 \pm 0.26$ MeV. In PDG~\cite{ParticleDataGroup:2022pth}, $\Xi_c(2815)^+$ is assigned as a $3/2^-$ state from the quark model expectation although the spin-parity has not been measured. Our calculation, assuming the $\lambda$-mode excitation, can roughly explain the total width, as shown in Table~\ref{tab:Xi_Q}. 
Interestingly, one can see that the contribution from the direct process here is of a similar order to that of $\Lambda_c(2625)$; however, its contribution is tiny compared to the $S$-wave decay to $\Xi_c^\prime(2645)\pi$.

The lower panel of Fig.~\ref{fig:Xi_c} presents the Dalitz and invariant mass plots for different decay modes. 
In the case of $\Xi_c(2815) \to \Xi_c^\prime(2645)\pi$, the decay occurs in an $S$ wave, and the Dalitz plots show a pronounced $\Xi_c^\prime(2645)$ resonance band with a very small nonresonant contribution. 
This situation is opposite to that of $\Lambda_c(2625)$, where the $\Sigma_c(2520)$ is not kinematically allowed, and the direct process has a considerable contribution.
Additionally, the $\Xi_c(2815)^+$ decays are intriguing as the Dalitz plot exhibits distinct structures of cross diagrams in Fig.~\ref{fig:diagram}, manifested as diagonal bands. 
For instance, for the $\Xi_c^+\pi^+\pi^-$ mode in Fig.~\ref{fig:Xi_c}~(d), only the $\Xi_c^+\pi^-$ combination can couple to $\Xi_c^\prime(2645)^0$, while $\Xi_c^+\pi^+$ cannot couple to any $\Xi_c^\prime(2645)$ state due to charge conservation.
For the $\Xi_c^+\pi^0\pi^0$ mode in Fig.~\ref{fig:Xi_c}~(e), the $\Xi_c^+\pi^0$ combination couples to $\Xi_c^\prime(2645)^+$, contributing equally to both the vertical and diagonal bands in the Dalitz plot. 
For the $\Xi_c^0\pi^+\pi^0$ in Fig.~\ref{fig:Xi_c}~(f), both $\Xi_c^\prime(2645)^0$ (vertical) and $\Xi_c^\prime(2645)^+$ (diagonal) appear simultaneously in the plot due to their different charges. However, the $\Xi_c(2815)^+\to\Xi_c^\prime(2645)^0\pi^+$ decay is favored, enhancing $\Xi_c^\prime(2645)^0$ band compared to $\Xi_c^\prime(2645)^+$ in the Dalitz plot. This distinct contribution of diagonal bands is clearly observed in the ${\Xi_c\pi}$ invariant mass distribution, although it has minimal impact on the ${\pi\pi}$ invariant mass distribution.

It should be noted that the coupling of the direct process is fixed according to the chiral-partner scheme~\cite{Kawakami:2018olq,Kawakami:2019hpp}, with the direct coupling set equal to the coupling at the second vertex in Fig~\ref{fig:diagram}. Therefore, verifying whether the nonresonant contribution is highly suppressed in this $\Xi_c(2815)^+$ decay would test the chiral partner structure in heavy baryons. Additionally, one can immediately see that the ${\pi\pi}$ invariant mass is flat, not showing an antisymmetric shape as in the case of $\Lambda_c(2625)^+$ decay. From this, we infer that the direct process is important when the resonance contribution is suppressed.

%--------------------------------------------------------
\subsection{$\Xi_b^*(1/2^-,3/2^-)$ decays} 
%--------------------------------------------------------

The low-lying $\Xi_b^*(1/2^-, 3/2^-)$ states have recently been identified by CMS~\cite{CMS:2021rvl} and LHCb~\cite{LHCb:2023zpu}. While both neutral states have been discovered, only the $\Xi_b(6100)^-$, corresponding to the $3/2^-$ state, has been observed, leaving the $\Xi_b^*(1/2^-)^-$ undetected. Therefore, we present the results for the neutral states, designating the observed $\Xi_b(6087)^0$ and $\Xi_b(6095)^0$ as $\Xi_b^*(1/2^-)$ and $\Xi_b^*(3/2^-)$, respectively.

%--------------------------------------------------------
\subsubsection{$\Xi_b(6087)^0$}
%--------------------------------------------------------

The $\Xi_b(6087)^0$ was recently discovered by LHCb~\cite{LHCb:2023zpu} via the $\Xi_b^{\prime -}(\Xi_b^0\pi^-)\pi^+$ decay. From its dominant decay via $\Xi_b^{\prime -}\pi^+$, it can be inferred that the state is most likely a $1/2^-$ state. 
Furthermore, the measured width of $2.43\pm 0.51\pm 0.10$ MeV is consistent with a $1/2^-$ state with a $\lambda$ mode, as shown in Table~\ref{tab:Xi_Q}. 
The two-pion emission  decay is primarily dominated by the $\Xi_b^{\prime}$ intermediate state. 
The decay via the $\Xi_b^{*\prime}$ intermediate state is negligible since it occurs virtually and through a $D$ wave. 

The Dalitz and invariant mass plots are shown in the upper panel of Fig.~\ref{fig:Xi_b}. In the decay mode (a) $\Xi_b^0 \pi^- \pi^+$, a simple structure with a single resonance band from $\Xi_b^{\prime -}$ is observed, while other modes show some kinematic reflections (diagonal bands). Specifically, for the $\Xi_b^0 \pi^0 \pi^0$ mode in (b), a smeared diagonal band is visible because the parent particle's width is only about 2 MeV, and the $\Xi_b^\prime$ resonance is small. Interestingly, Fig.~\ref{fig:Xi_b} (c) only shows the $\Xi_b^{\prime-}$ resonance; the $\Xi_b^{\prime 0}$ is absent as it lies outside the plot range discussed in Fig.~\ref{fig:boundary} because $\Xi_b^{\prime 0}$ cannot decay into $\Xi_b^- \pi^+$ due to limited phase space. Consequently, it results in a very small partial decay width in Table~\ref{tab:Xi_Q}. It is important to note that $\Xi_b^{\prime 0}$ has not yet been observed experimentally as summarized in Table~\ref{tab:mass}, and its mass and width are assumed to be the same as $\Xi_b^{\prime -}$ in this study. Additionally, the direct process is hardly visible, overshadowed by the dominant resonant contribution. This finding is consistent with those in Refs.~\cite{Kawakami:2018olq,Kawakami:2019hpp}.

%--------------------------------------------------------
\subsubsection{$\Xi_b(6095)^0$}
%--------------------------------------------------------

The $\Xi_b(6095)^0$ was recently discovered by LHCb~\cite{LHCb:2023zpu} via the $\Xi_b^{*\prime -}(\Xi_b^0\pi^-)\pi^+$ decay and is possibly an isospin partner of $\Xi_b(6100)^-$ found earlier by CMS~\cite{CMS:2021rvl}. From its dominant decay via $\Xi_b^{*\prime -}\pi^+$, it is consistent with a $3/2^-$ state. Furthermore, the measured width of $0.50$ MeV agrees with a $3/2^-$ state with a $\lambda$-mode excitation, as shown in Table~\ref{tab:Xi_Q}. The decay is dominated by the $S$-wave $\Xi_b^{*\prime}\pi$ mode, which subsequently decays into $\Xi_b\pi\pi$.

The Dalitz and invariant mass plots are displayed in the lower panel of Fig.~\ref{fig:Xi_b}. The discovery decay mode $\Xi_b^0\pi^-\pi^+$ in (a) exhibits a dominant $\Xi_b^{*\prime}(3/2^+)$ resonance band near the boundary. The $\Xi_b^{\prime}(1/2^+)$ resonance lies inside the Dalitz plot, but the contribution is relatively small due to its $D$-wave nature. 
A similar situation applies to decay modes in (e) and (f), where the $\Xi_b^{\prime}$ resonance is hindered by the $S$-wave resonance. Additionally, the $\Xi_b^0\pi^0\pi^0$ in (e), where the phase space is slightly larger, the $\Xi_b^{*\prime}$ resonance band now resides well inside the plot. This phenomenon is reminiscent of that observed in the $\Lambda_c(2595)$ decay shown in Fig.~\ref{fig:Lc_dalitz12}.
Once again, the direct process is found to be negligibly small despite the narrow width of $0.5$ MeV, hindered by the dominant $S$-wave decay, as in previous cases.
Furthermore, the smearing effect is also negligible due to its narrow width.

%--------------------------------------------------------
\section{Conclusion}
%--------------------------------------------------------
\label{sec:conclusion}

We have analyzed the two-pion emission decays of the singly charmed and bottom baryons $\Lambda_Q$ and $\Xi_Q$ with $J^P = 1/2^-$ and $3/2^-$, which belong to the antisymmetric flavor triplet $\bar{\bm{3}}_F$, as presented in Table~\ref{tab:mass}. 
Thanks to recent experimental development, most of these baryons have already been discovered. 
In the present work, we consider both sequential processes through intermediate states of the symmetric flavor sextet $\bar{\bm{6}}_F$, $\Sigma_Q$ and $\Xi_Q^\prime$, with $J^P = 1/2^+$ and $3/2^+$. 
The direct two-pion process is also included, which is important for comparison with experimental data. 
The coupling for the sequential process is derived from the chiral quark model~\cite{Arifi:2017sac,Arifi:2018yhr}, while the direct process is assumed to follow the chiral partner scheme~\cite{Kawakami:2018olq,Kawakami:2019hpp}. 
The convolution of the initial parent's mass for the Dalitz plot is also considered to obtain a more realistic comparison with the data.

Based on the comparison with the available experimental data, our decay analysis provides further support that these resonances are consistent with a heavy quark symmetry (HQS) doublet with $(1/2^-, 3/2^-)$ and brown-muck spin $j=1$ in the $\lambda$-mode excitation~\cite{Nagahiro:2016nsx,Yoshida:2015tia}. 
Our study includes not only the total decay width and branching fractions but also a systematic analysis of the Dalitz and invariant mass plots. 
In particular, we compared our quark model results with recent Belle data~\cite{Belle:2022voy} for the $\Lambda_c(2625)^+$ and found a good description of the Dalitz plot as well as the ${\Lambda_c^+\pi^+}$ and ${\pi^+\pi^-}$ invariant mass distributions. 
The ratio of the $\Sigma_c(1/2^+)$ and the shoulder of $\Sigma_c^*(3/2^+)$, which is reflected in the ${\Lambda_c^+\pi^+}$ distribution, can be well described using the $\lambda$ mode. 
This ratio would be poorly described if assigned to the $\rho$ mode~\cite{Arifi:2018yhr}. 
Moreover, the chiral-partner scheme effectively estimated the direct two-pion coupling process, explaining the asymmetry in the ${\pi^+\pi^-}$ invariant mass distributions. 
We also compared our results with some LHCb data~\cite{LHCb:2012kxf} for the $\Lambda_b(5920)^0$ decay; however, more data is needed to confirm the observed asymmetry.

For decays involving $\Xi_Q$ baryons, the direct process is suppressed when the $S$-wave resonance contributes, resulting in no visible asymmetry in the ${\pi^+\pi^-}$ invariant mass distributions. 
Only $\Xi_c(2790)^+$ decay shows a dominant direct process as seen in Fig.~\ref{fig:Xi_c}.
The contribution of this nonresonant process predicted by the chiral-partner structure in heavy baryon sectors may provide a further test that can be verified in experiments.
For example, the $\Xi_b(6095)^0$ with a narrow width of 0.5 MeV has large resonance contribution but has negligible direct process.
This situation is opposite to the $\Lambda_c(2625)^+$ with similar decay width. 
Additionally, the $\Xi_b(6087)^0$ and $\Xi_b(6095)^0$ decays may exhibit an isospin-breaking effect similar to that observed in the $\Lambda_c(2595)^+$.

Further verification by LHCb, BelleII, and BESIII experiments will be essential for defining the structure of negative parity heavy baryons. This is particularly important because some resonances are near some thresholds, making it crucial to clarify their nature through the quark model and to identify any exotic behaviors.

%--------------------------------------------------------
\section*{Acknowledgements}
%--------------------------------------------------------
We thanks Makoto Oka and Atsushi Hosaka for useful discussions. 
N.P. acknowledges the hospitality from RIKEN during her stay for a completion of this work. N.P. is also financially supported by the National Astronomical Research Institute of Thailand (NARIT). A.J.A. was supported by RIKEN special postdoctoral researcher program. D.S. is supported by the Fundamental Fund of Khon Kaen University and has received funding support from the National Science, Research and Innovation Fund and supported by Thailand NSRF via PMU-B [grant number B37G660013]. D.S. is also financially supported by the Mid-Career Research Grant 2021 from National Research Council of Thailand under a Contract no. N41A640145. 

%--------------------------------------------------------
\bibliography{reference}
%--------------------------------------------------------

\end{document}